\documentclass[preprint, times]{elsarticle}   %
\usepackage[utf8]{inputenc}

\usepackage{amsfonts,amsmath,amssymb}
\usepackage{graphicx}
\usepackage{color,soul}
\usepackage{algorithm2e}
\usepackage{url}
\usepackage{textcomp}
\usepackage{array}
\usepackage{todonotes}
\usepackage{hyperref}
\usepackage{units}
\usepackage{lineno}[displaymath, mathlines]
\usepackage[skip=0pt]{subcaption}
\graphicspath{{./}{./res/}}
\newcommand{\q}[1]{``#1''}
\newcommand{\hmin}{h_{\text{min}}}
\newcommand{\hmax}{h_{\text{max}}}
\newcommand{\hm}{h_{\text{s}}}

\newcommand{\subframe}[1]{#1}

\makeatletter
\define@key{Gin}{domainSubfig}[]{\setkeys{Gin}{trim=11mm 0mm 13mm 2mm,clip,width=\textwidth}}
\define@key{Gin}{domainHalfFig3d}[]{\setkeys{Gin}{trim=3mm 0mm 5mm 0mm,clip,width=\textwidth}}
\define@key{Gin}{domainHalfFig}[]{\setkeys{Gin}{trim=11mm 1mm 13mm 2mm,clip,width=0.47\textwidth}}
\define@key{Gin}{fullFig}[]{\setkeys{Gin}{width=0.95\textwidth}}
\define@key{Gin}{halfFig}[]{\setkeys{Gin}{width=0.48\textwidth}}
\define@key{Gin}{oneThirdFig}[]{\setkeys{Gin}{width=0.32\textwidth}}
\makeatother

\newfont{\ssdcmm}{cmbx10 scaled \magstep1}
\newfont{\ssdcm}{cmssdc10 scaled \magstep2}
\newfont{\ssdcs}{cmssdc10 scaled \magstep3}
\newfont{\ssdcv}{cmssdc10 scaled \magstep4}

\journal{Computer Physics Communications}

\begin{document}

\begin{frontmatter}

	\title{Parallel domain discretization algorithm for RBF-FD and other
	meshless
		numerical methods for solving PDEs}
	\author[1]{Matjaž Depolli\corref{cor1}}
	\ead      {matjaz.depolli@ijs.si}
	\author[1]{Jure Slak}
	\ead      {jure.slak@ijs.si}
	\author[1]{Gregor Kosec}
	\ead      {gregor.kosec@ijs.si}

	\address[1]    {Jožef Stefan Institute, Jamova cesta 39, SI-1000 Ljubljana,
	Slovenia}
	\cortext[cor1] {Corresponding author}

\begin{abstract}
	In this paper, we present a novel parallel dimension-independent node positioning algorithm that is capable of generating nodes with variable density, suitable for meshless numerical analysis. A very efficient sequential algorithm based on Poisson disc sampling is parallelized for use on shared-memory computers, such as the modern workstations with multi-core processors. The parallel algorithm uses a global spatial indexing method with its data divided into two levels, which allows for an efficient multi-threaded implementation. The addition of bootstrapping enables the algorithm to use any number of parallel threads while remaining as general as its sequential variant. 
	We demonstrate the algorithm performance on six complex 2- and 3-dimensional domains, which are either of non rectangular shape or have varying nodal spacing or both. We perform a run-time analysis of the algorithm, to demonstrate its ability to reach high speedups regardless of the domain and to show how well it scales on the experimental hardware with 16 processor cores. We also analyse the algorithm in terms of the effects of domain shape, quality of point placement, and various parallelization overheads.
\end{abstract}

\begin{keyword}
	parallel \sep
	point fill \sep
	meshless \sep
	Poisson disc sampling \sep
	performance
\end{keyword}

\end{frontmatter}

\section{Introduction}
Many natural phenomena can be described with partial differential equations
(PDEs) that are in most practical cases unsolvable in closed form and numerical
approaches are used instead. An important part of the whole numerical solution
procedure is domain discretisation, i.e.\ a division of the domain of interest
into small discrete parts. The simplest way to numerically address PDEs is to
use the Finite Difference Method (FDM) where the domain is
covered with equidistant orthogonal mesh that can be trivially constructed.
Despite its simplicity, FDM can be a powerful tool for
simple geometries; however, complications arise with irregular geometries or
when performing adaptive refinement. A more advanced method is the
Finite element method (FEM), a standard numerical tool for engineering and
scientific simulations that solves PDEs in weak form. Although FEM is capable
of solving PDEs on irregular domains and supports all types of adaptivity,
meshing of realistic 3D domains itself remains a problem that
often requires user assistance and thus cannot be fully
automated~\cite{trobec2015parallel}.

As a response to geometrical limitations of simple methods like FDM and the
complexity of meshing required by FEM, a new
class of meshless or often also referred to as meshfree methods
emerged~\cite{liu2002mesh}. The
difference between the mesh-based and the meshless methods is in consideration
of the relations between nodes. The mesh-based methods structure
nodes into polygons that cover entirely the domain of interest, the process
referred to as meshing. On the other hand, the meshless methods fully define
the relation between nodes only by internodal distances.
An important implication of this distinction is that the meshless methods
can solve PDEs on a set of scattered nodes, without a
mesh~\cite{trobec2015parallel, liu2002mesh}. The recent review on Taylor series
expansion based methods discusses also the performance of strong form meshless
methods in problems with non-smooth solutions~\cite{Jacquemin2020}.

In the early stages of meshless development, some authors communicated that
completely arbitrary nodes, e.g. randomly generated, can be used in meshless
methods~\cite{liu2002mesh}, making node
generation seemingly trivial. Today, it is widely accepted that despite
meshless generality regarding the node positions, nodes have to be positioned
with certain rules in mind~\cite{SlakKosec2019NodeGen}. Despite the limitations
in
node positioning, the discretisation of
the domain with scattered nodes is still a less complex operation than
meshing~\cite{SlakKosec2019NodeGen, zienkiewicz2005finite}.

In the last few years, a substantial effort has been put into the development of
algorithms dedicated to the meshless discretisation, i.e.\ the node
positioning algorithms. In general, there are different node positioning
strategies ranging from expensive iterative relaxation
algorithms~\cite{hardin2004discretizing, kosec2018local, liu2010node},
advancing
fronts~\cite{lohner2004general}, to sphere packing
algorithms~\cite{li2000point}.
Recently, the Poisson disc
sampling (PDS)~\cite{cook1986stochastic} based positioning algorithm that
generate nodes suitable for a meshless numerical analysis has been
proposed~\cite{SlakKosec2019NodeGen, shankar2018robust}.
In~\cite{SlakKosec2019NodeGen} the authors
introduced the PDS based algorithm that places nodes with spatially variable
nodal density in an arbitrary domain in
two, three or more dimensions. The quality of generated nodes has been, after
a thorough analyses, verified by solving
a transient solution of a coupled heat and momentum transport (natural convection
problem) in a complex 3D domain with Radial Basis Functions Generated Finite
Difference (RBF-FD)
method~\cite{fornberg2015primer}. 

Encouraged by the results presented in~\cite{SlakKosec2019NodeGen, slak2018adaptive} the PDS based algorithm
was promoted to a default
discretisation algorithm for an open source meshless project
Medusa~\cite{medusa}. Since its integration in Medusa, it has been used
in high order solution of Poisson's equation, including a
solution of a problem in 4D~\cite{janvcivc2019analysis}, and in an engineering
simulations of
overhead
power thermal rating~\cite{maksic2019cooling}. The algorithm has also been
generalised to
the discretisation of $n$-D parametric surfaces, an important step towards the
coupling of meshless methods and Computer-Aided Design
(CAD)~\cite{duh2020fast}. Generating nodes on
parametric surfaces could also be used in conjunction with a dedicated surface
reconstruction algorithm~\cite{drake2021implicit} to numerically address
problems on or within the
geometry given by a real point cloud data, for which there is no guarantee
that the points are distributed according to the requirements for
meshless nodes.

The algorithm~\cite{SlakKosec2019NodeGen} inherently supports $h$-adaptive
numerical analysis through the spatially variable node density that directly 
affects the internodal distance $h$. In adaptive approach, nodal density 
function is constructed in such way that in areas with higher expected error 
$h$ is reduced and in areas with lower expected error $h$ is increased. The 
crucial part in constructing appropriate nodal density function is  
error estimate that defines the refinement criteria. Error estimates 
have been in the meshless context
researched to some degree both in weak and strong form solutions of elasticity
problems~\cite{Ebrahimnejad,Angulo}. In the
context of RBF-FD method a ZZ~\cite{Zienkiewicz_1} type error indicator has
been demonstrated in~\cite{Oanh}, and in meshless methods relying on the
least-squares approximation
residual based error indicator have been discussed in~\cite{Sang}. Recently, 
also a Partition-of-unity based error indicator has been demonstrated in 
adaptive RBF-FD solution of Poisson's equation~\cite{mipro2021}. Besides 
indicators that extract knowledge about the error solely from a numerical
method, ad-hoc error indicators that interpret the physical solution have been 
reported in~\cite{Davydov, kos2}. The latter approach has been also
implemented with the discretization algorithm at hand in a fully adaptive
solution of elasticity problem~\cite{slak2018adaptive}, where the nodal density 
function has been constructed on top of data provided by ad-hoc error 
indicator. 

Another important topic, especially in context of adaptive solution, is the 
selection of appropriate stencils from the point clouds. In context of RBF-FD 
the topic has been thoroughly investigated in adaptive solution of Poisson's 
equation~\cite{Davydov}, including the problems with point 
singularities~\cite{Oanh}.
In~\cite{Jacquemin2021}, the authors present a method for stencil selection, based on 
the visibility criterion, for convex, concave, and singular problems in 2D and 3D. 
In~\cite{Davydov, Oanh} authors demonstrated 
effective stencils of $6$ or $5$ nearby points that are sufficiently uniformly 
distributed around central node to support RBF-FD, and successfully solved 
several test problems. Nevertheless, often a much simpler approach where a 
certain, e.g. $12$ or more in 2D~\cite{janvcivc2019analysis}, number of closest 
nodes form a stencil is 
used. Although such approach is easy to implement and generalise to higher 
dimensions, it is worth noting that it is also 
computationally less effective, since more supporting nodes are required 
for stable computation, especially in adaptive approach.  

Although the algorithm presented in~\cite{SlakKosec2019NodeGen} is
computationally
effective, it can become a computational bottleneck, especially as it cannot be
executed in parallel, contrary to most of the other parts of the
meshless solution procedure. It is therefore the goal of this paper to
discuss the development of the parallel version of the PDS positioning
algorithm presented in~\cite{SlakKosec2019NodeGen}.

The PDS algorithm itself has been done in parallel before.
In \cite{wei08parallelpoisson}, where the authors partition the domain into
independent cells and then groups cells in such a way that all the cells from
the same group are sufficiently far apart to be sampled in parallel,
without any conflicts.
The algorithm trades some of the performance for the emphasis on the blue
noise property of samples, a property not needed in the scope of meshless
analysis.
Although it supports non-rectangular domains and adaptive sampling, it does so
on account of performance. The authors admit that the non-uniform sampling can
be incredibly inefficient.
In \cite{ying2013intrinsic}, PDS is parallelised, with the emphasis on blue
noise consistency.
The authors use the dart-throwing method to generate a large set of points that
they prune later.
Both dart-throwing and pruning can be done in parallel.
Parallel pruning uses a location independent priority value given to every
generated point to resolve the conflicts between concurrent processes.
The performance is not satisfactory, though, since dart throwing is not
efficient for non-rectangular domains, and the generation of a large number of
additional points increases memory requirements and processing overheads.

In this paper, we attack the problem differently to the aforementioned parallel
attempts, since except for the parallelization, our goals are different.
The goals are identical to the ones posed to the sequential algorithm in
\cite{SlakKosec2019NodeGen}, namely a fast execution and enforced proper
inter-nodal spacing, while the distribution of placed nodes is free to
diverge from the blue noise pattern.
We developed a parallel algorithm based on an advancing-front method,
inherited from its sequential parent, which does not suffer performance
penalties on non-rectangular domains and non-uniform sampling.
However, parallelization options are limited for advancing-front algorithms,
since addition of new points is constricted to areas covered by the open
fronts. Therefore in the proposed algorithm, the main novelties are the
partitioning step that enables creation of multiple fronts and the dynamic
parallel spatial indexing that supports parallel manipulation of fronts.

The rest of the paper is organised as follows: in
section~\ref{sec:sequentialAlgorithm}
the sequential algorithm and its application in RBF-FD is discussed,
in section~\ref{sec:parallel_algorithm} a parallel algorithm is presented,
in section~\ref{sec:experiments} experiments with the newly developed algorithm
are
given, followed by discussion in section~\ref{sec:discussion}; in the last
section~\ref{sec:conclusion}, conclusions and guidelines for future work are
given.

\section{Sequential algorithm}
\label{sec:sequentialAlgorithm}

The algorithm operates on \q{points} in $d$-dimensional space, which are only
referenced as \q{nodes} in the context of a numerical method.
A domain $\Omega \subset \mathbb R^d$, a spacing function $h$, and $n_s$
\q{seed points} $p_1, \ldots, p_{n_s}$ are taken as input parameters. The
domain is assumed to be bounded and is represented with its characteristic
function $\chi_\Omega$,
\begin{equation}
\chi_\Omega(p) = \begin{cases}
1, p \in \Omega \\
0, p \notin \Omega.
\end{cases}
\end{equation}
The spacing function $h\colon\Omega\to (0, \infty)$ is assumed to be bounded
away from zero, which ensures the finiteness of the algorithm. At least one seed
point is assumed to be given for each connected component of $\Omega$ and all
seed points are assumed to lie inside of $\Omega$. Apart from that, there are no
additional assumptions needed on the topological properties of $\Omega$ nor on
the function $h$ for the algorithm to work. Despite that, the algorithm is
expected to work best if we also assume that the values of the spacing function
$h$ are at least an
order of magnitude smaller than the sizes of geometric features of $\Omega$,
such as narrow choke points. This assumption is natural from the point of view
of PDE discretisations, as any geometric feature must be properly
discretised to solve the eventual problem. Similarly, the spacing
function $h$ is usually continuous, or even Lipschitz continuous, to produce
gradual nodal spacing changes.

The algorithm works as an advancing-front algorithm and places points in a
breadth-first-search manner, starting from the seed points. These are initially
put in a queue, and the algorithm considers the points from the queue (also
called \q{active} points) until the queue
is empty. Besides the queue, the seed points are also inserted into a spatial
index $S$. In each iteration, a point $p$ is taken from the queue and expanded.
That means that new
candidate points $\{c_j\}_{j=1}^{n_{ct}}$ are generated from it, by placing
them uniformly on a $d$-dimensional sphere centred at $p$ with radius $h(p)$.
The newly generated
candidates are
then processed in sequence: if a candidate $c$ lies outside $\Omega$ it is
immediately discarded. Likewise, if $c$ lies too close to any of the existing
points, it is also discarded. This proximity check is performed by finding the
distance to the closest neighbour of $c$ using the spatial index $S$, and
checking that the distance is smaller than $h(c)$. If the candidate $c$ is not
discarded by any of the tests, it is accepted, inserted into the queue and into
the index $S$. The algorithm ends when the queue empties, and all the generated
points are then contained in the index $S$.

The generation of candidates is done using the recursive discretisation of the
$d$-dimensional sphere along the last spherical coordinate. The number of
candidates is
controlled using the parameter $n_c$, which represents the number of generated
candidates on the great circle, and the total number of candidates
$n_{ct}$ is in the order of $O(n_c^{d-1})$, falling to the curse of
dimensionality. The higher
the number of candidates, the better the quality of the final discretisation.
However, increasing the number of candidates comes with diminishing returns,
and setting $n_c$ between 10 and 30 is usually sufficient.
Further details about the candidate generation
and the sequential algorithm are available in the
paper describing the sequential algorithm~\cite{SlakKosec2019NodeGen}.

The time and space complexity of the sequential algorithm are
best described in an
output sensitive way, in terms of the total number of points $n_p$.
The algorithm performs at most $n_{ct}$ evaluations of $\chi_\Omega$ and
closest neighbour searches in $S$. Additionally, each point is inserted once,
totalling $n_p$ insertions in $S$, and the rest of the operations are
(amortized) constant. The time complexity is thus
\begin{equation}
T_{\text{sequential}} = O(n_p T_I + n_{ct} n_p (T_\Omega + T_Q)),
\end{equation}
where $T_Q$ and $T_I$ are time complexities of query and insertion operations
in S, respectively, and $T_\Omega$ is the time complexity of evaluation of
$\chi_\Omega$.
For a practical implementation, it is usually
assumed that $T_\Omega$ is constant, and $T_I$ or $T_Q$ are on average
logarithmic in the number of points, such as when using a $k$-d tree for $S$.
Therefore, the asymptotic time complexity of sequential algorithm can be
simplified to:
\begin{equation}
\label{eq:complexity_sequential}
T_{\text{sequential}} = O(n_{ct} n_p \log(n_p)),
\end{equation}
The space complexity is bounded by how much space is needed in $S$ to store
$n_p$ points.
It depends on the spatial indexing used, but in case of $k$-d trees it is
\begin{equation}
S_{\text{sequential}} = O(n_p).
\end{equation}

\begin{figure}[ht]
	\includegraphics[oneThirdFig]{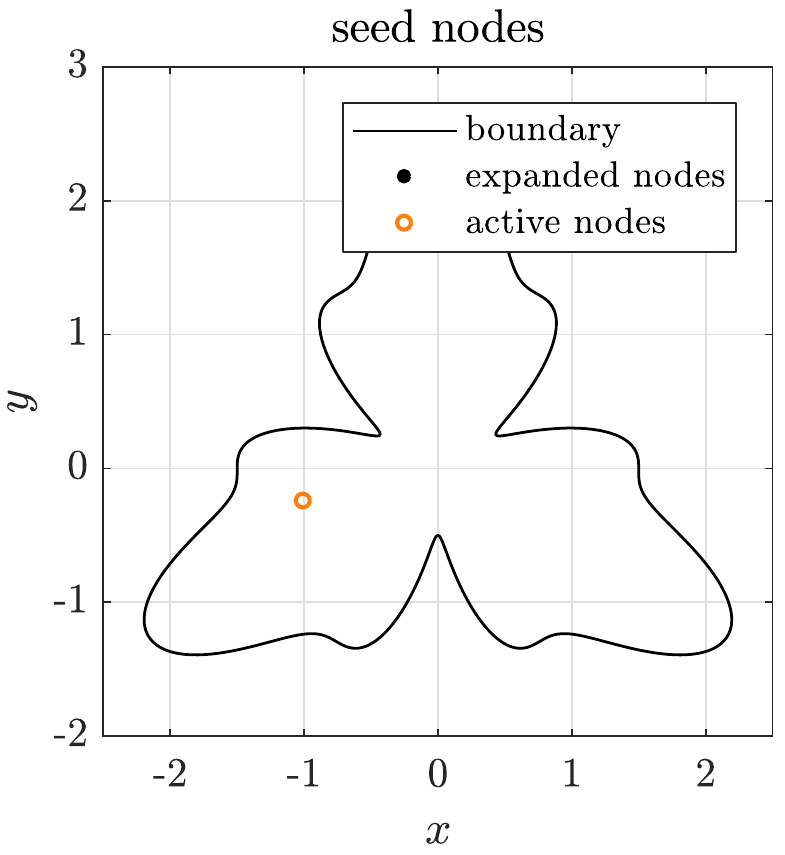}
	\includegraphics[oneThirdFig]{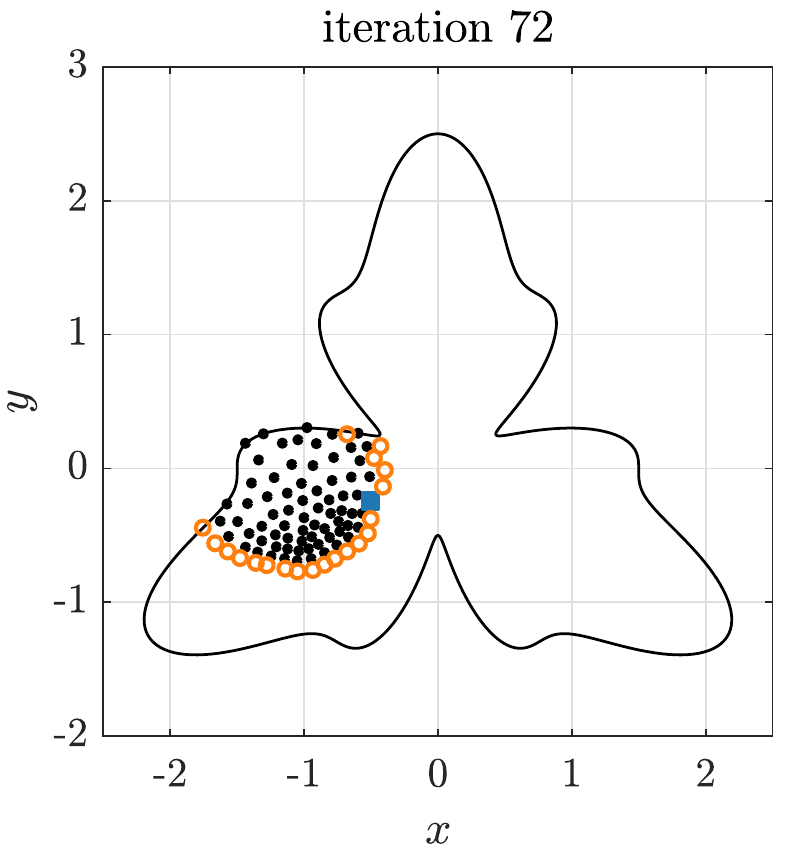}
	\includegraphics[oneThirdFig]{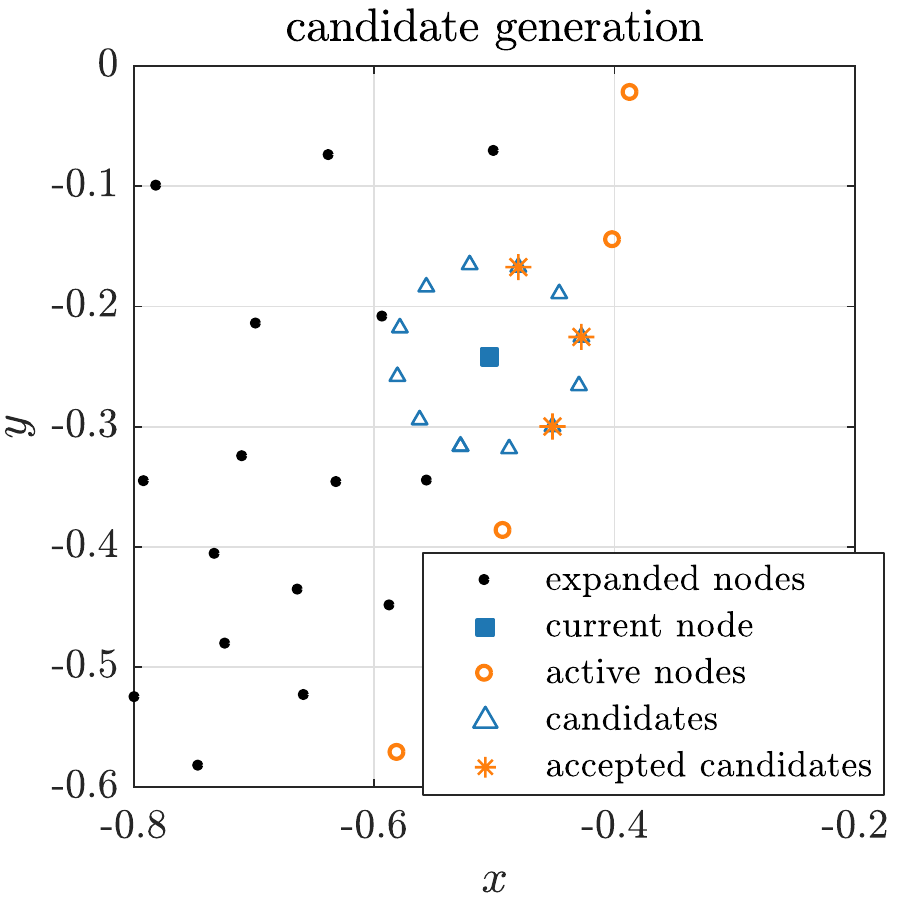}
	\includegraphics[oneThirdFig]{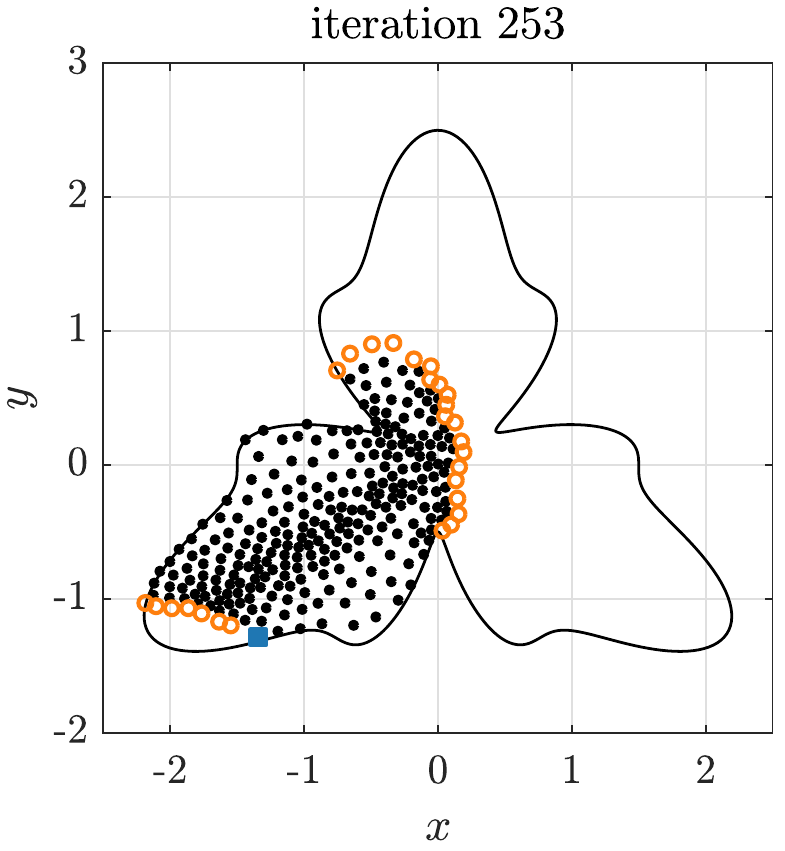}
	\includegraphics[oneThirdFig]{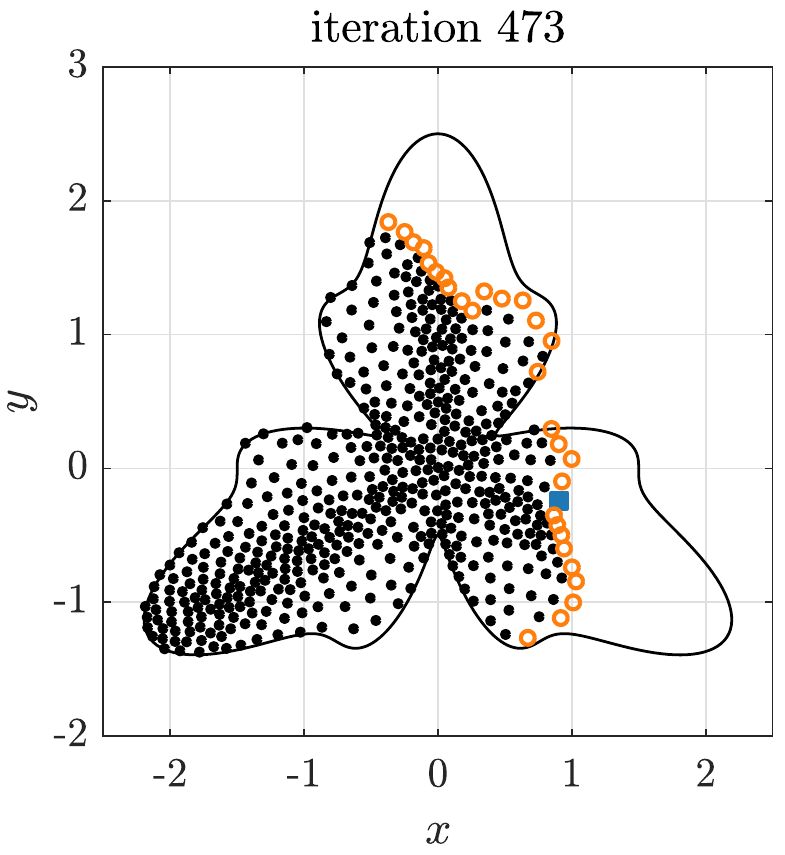}
	\includegraphics[oneThirdFig]{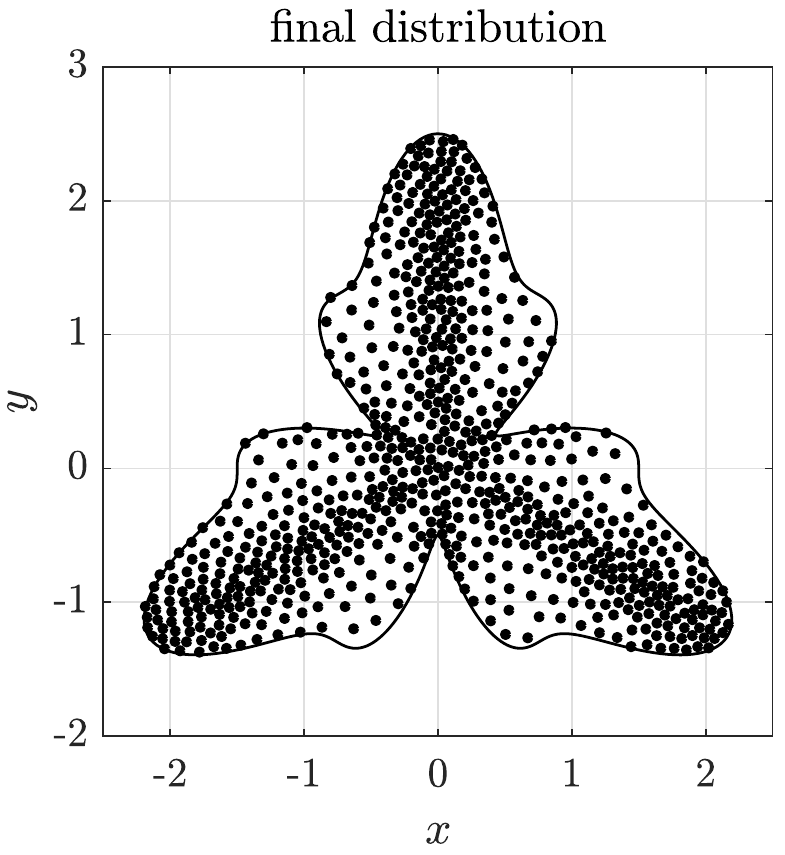}
	\caption{Execution of the point generation algorithm on the "clover"
		test domain, defined by $\Omega_t$, with
		spacing function $h_t$ and $n_c = 12$. The top right figure shows the
		candidate generation and selection around the \q{current point} in
		iteration 72.}
	\label{fig:seq-demo}
\end{figure}

Figure~\ref{fig:seq-demo} shows a sample execution of the algorithm.
The test domain, which we shall refer to as \q{clover}, is the area
enclosed by the polar curve
\begin{equation}
r(\theta) =  3/2 - \cos^3(3(\theta-\pi/6)),
\end{equation}
i.e.
\begin{equation}  \label{eq:clover}
\Omega_t = \{(x, y); \; x^2+y^2 < r_p^2(\arg(x, y)) \},
\end{equation}
where $\arg(x, y)$ represents the angle that the ray from the origin through
point $p=(x, y)$ makes with the positive part of the $x$ axis, or as commonly
computed
by the \verb|atan2| function. For demonstration we use spacing function
\begin{align} \label{eq:clover-den}
h_t(x, y) &= \hmin + (\hmax-\hmin) h_{\text{base}}(x, y), \\
h_{\text{base}}(x, y) &= \cos^2(3\arg(x, y))
\tanh(\sqrt{x^2+y^2}),
\end{align}
where $h$ is defined in such a way that it varies between the minimal spacing
$\hmin$ and maximal spacing $\hmax$.
The same test domain $\Omega_t$ and
the spacing function $h_t$ will be used throughout the paper.
The number of placed points will be controlled by varying $\hmin$ and $\hmax$,
but their ratio will be set to constant: ${\hmax}/{\hmin}=5$, so that it can be
visualised while being sufficiently complex to make the placement
problem non-trivial.
This domain showcases both irregularity and spacing variability, which can
cause point placing algorithms to run inefficiently.
At the same time, the domain is two-dimensional, which makes the visualisations
much clearer.
However, several three-dimensional examples are also included later in the
paper.

\subsection{Solving a PDE on scattered nodes}
\label{sec:solving_pde}
Although we focus on RBF-FD in this paper, the nodes generated with the
presented algorithm can be used by any method that can operate on scattered
nodes, such as Element-Free Galerkin (EFG)~\cite{li2021linearized} or
MLPG~\cite{trobec2015parallel}.
In our implementation of RBF-FD~\cite{slak2021medusa}, the method expects the
domain discretisation of a domain $\Omega$ in the form of a list of nodes $X =
{x_1,
	\ldots, x_N}$, some of which are in the interior, and some lie on the
	boundary
of $\Omega$. Each node $x_i$ also has a list $I(x_i)$ of indices of its
neighbours. The neighbours of each node constitute its stencil, and the
stencils are usually computed as a part of the solution procedure. This is most
often done by selecting some fixed number $n$ of closest neighbours for each
$x_i$ (by Euclidean distance) and assigning their indices to $I(x_i)$.
While node positioning on the boundary and in the interior is done with a
dedicated algorithms, stencil selection is often a simple query operation on
the spatial index.

\begin{figure}[ht]
	\centering
	\includegraphics[height=6cm]{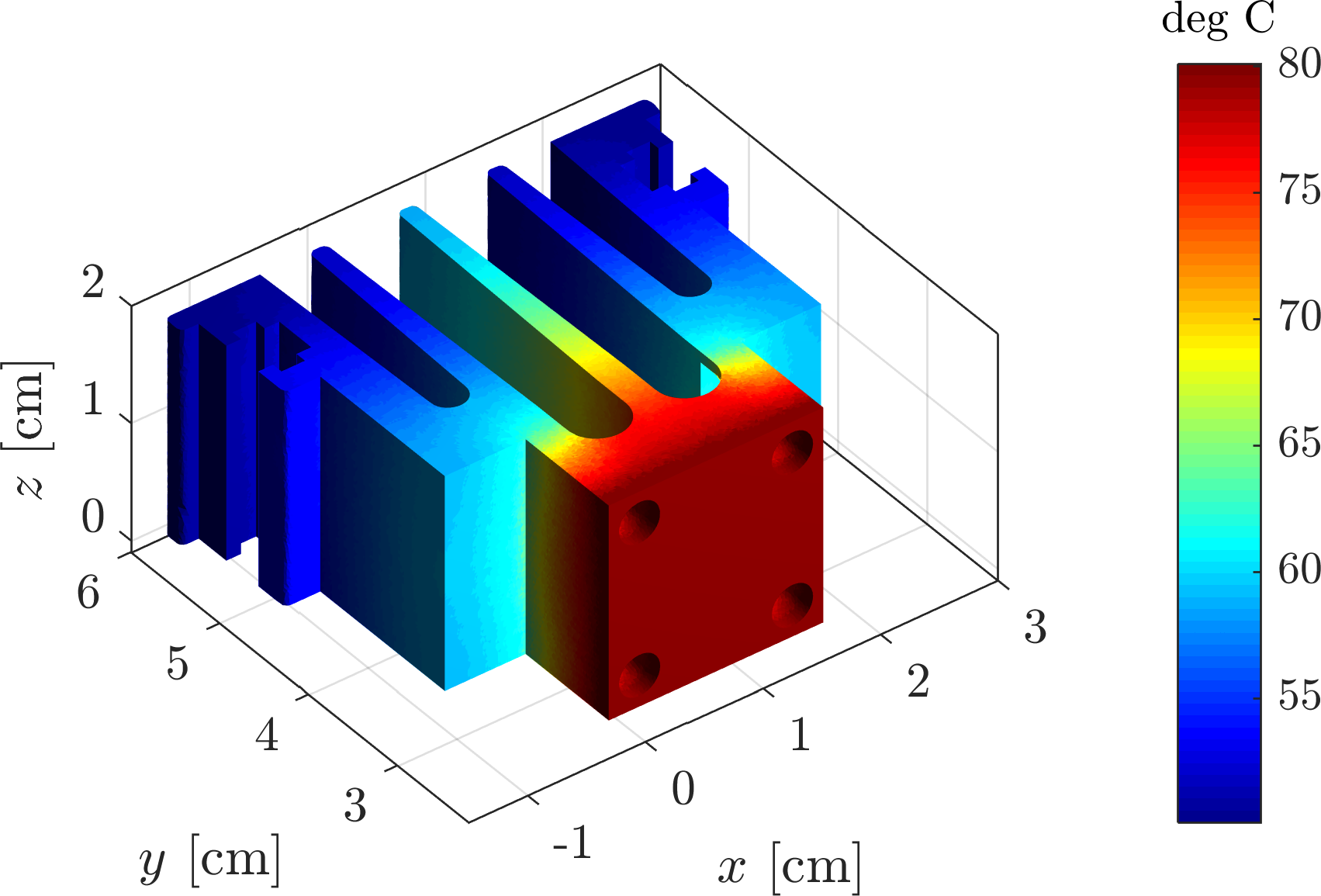}
	\caption{Temperature distribution in a heatsink, obtained by
	solving~\eqref{eq:pois-start}~through~\eqref{eq:pois-end} using RBF-FD with
	sequential
		execution.}
	\label{fig:heatisnk}
\end{figure}

As an example, we consider the following boundary value problem
\begin{align} \label{eq:pois-start}
\nabla^2 u &= 0  \text{ in } \Omega \\
u + \alpha \frac{\partial u}{\partial \vec{n}} &= u_a \text{ on } \Gamma_r \\
u &= u_b \text{ on } \Gamma_d
\label{eq:pois-end}
\end{align}
that represents a steady-state temperature distribution in domain $\Omega$,
shown in Figure~\ref{fig:heatisnk}. The domain $\Omega$ represents a heatsink,
obtained from~\cite{heatsink}, and
$\Gamma_d$ denotes the flat surface at the bottom of heatsink, at
roughly $y = \unit[2.5]{cm}$, and
$\Gamma_r$ is the rest of the domain's surface.
The surface $\Gamma_d$ is heated to a constant temperature
of $u_b = \unit[80]{^\circ C}$, and the rest of the surface experiences a flux
proportional to the difference between the surface temperature and the ambient
temperature $u_a = \unit[20]{^\circ C}$, with the proportionality constant
$\alpha = \unit[200]{cm}$.

One of the solutions $u$ is shown in Figure~\ref{fig:heatisnk}.
This particular solution was obtained using $N = 214458$ discretisation nodes,
of which 176896
were in the interior. Stencils of $n = 35$ closest neighbours were used, as
well as
Polyharmonic RBF $\phi(r) = r^3$ and augmentation with monomials up to 2nd
order. Ghost nodes were used to handle Robin boundary conditions.

\begin{figure}[ht]
	\centering
	\includegraphics[height=4.75cm]{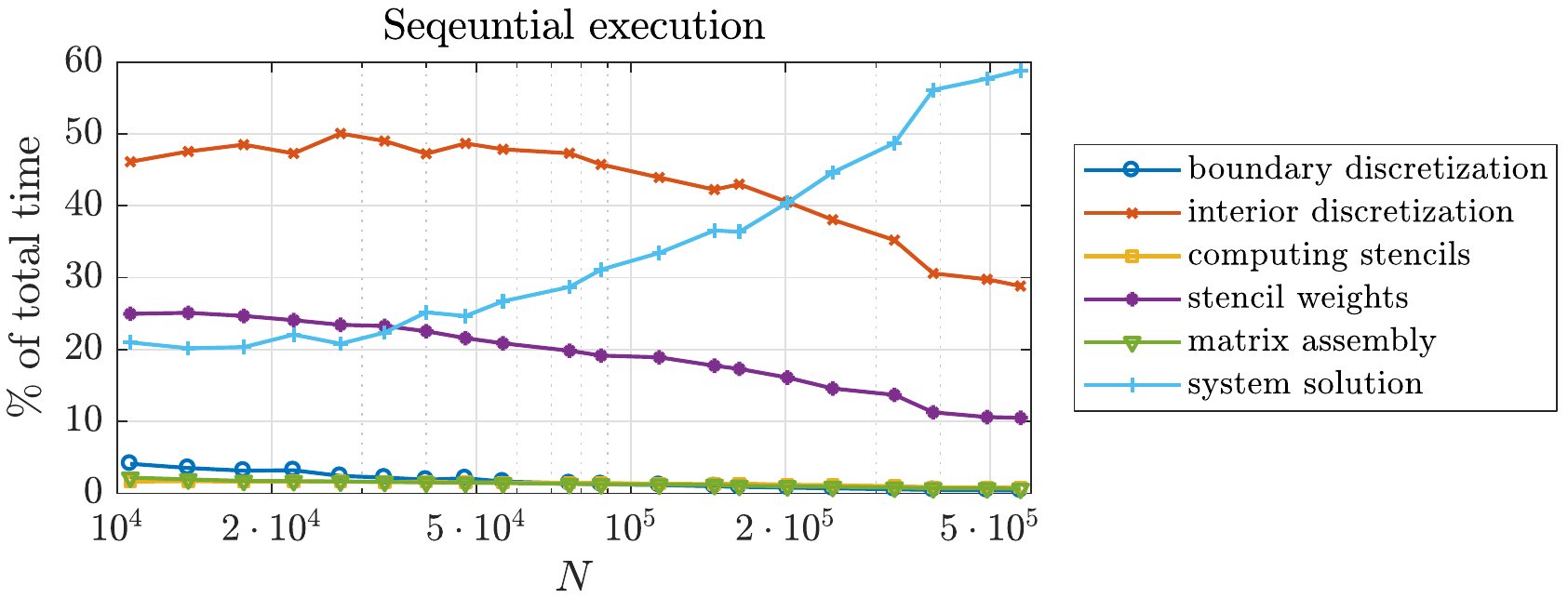}
	\caption{Distribution of running time between the parts of the
		solution procedure.}
	\label{fig:heatsink-time-distr}
\end{figure}

To determine the parallelisation opportunities, we separately measure each
step's execution time of the solution procedure. These steps include
placing the points on the boundary, placing the points in the interior,
computing stencil nodes,
computing stencil weights, assembling the global linear system, and solving it.
Figure~\ref{fig:heatsink-time-distr} shows the proportion of the total running
time that each of the parts takes for discretisations of different densities.

The most time-consuming parts of the solution procedure are the solution of the
final system, the computation of stencil weights, and the point placing in the
interior. All of these are worth parallelising. In fact, computing stencil
weights can be trivially done in parallel, as computations for each point are
independent. Parallelisation of linear system solvers is a whole research field
of its own, with several implementations readily available online. The final
part is
the parallelisation of the node positioning algorithm, which is the focus of
this
paper.

\section{Parallel algorithm}
\label{sec:parallel_algorithm}

The development of a parallel algorithm started from the sequential version.
We first analysed the sequential algorithm and found the following:
\begin{enumerate}
	\item There are no parallel opportunities worth exploring within a single
	point expansion.
	\item Expansions are prospective as units of parallel work.
	\item Expansions of points are local in the sense that they do not require
	knowledge of the whole global state, i.e.\ all of the placed points, but
	only
	their immediate neighbourhood.
	\item \label{finding:distant_points} Expansions of $p_1$ and $p_2$ can be
	done concurrently if points are distant enough, i.e., distance between
	points $p_1$ and $p_2$ exceeds max($h(p_1),h(p_2)$).
\end{enumerate}

We use these findings to design the parallel algorithm implemented as a set of
sequential algorithms working in parallel.
These sequential algorithms are transformed into threads, running on a single
computer, sharing a global spatial indexing structure.
For the threads to run independently, they should operate on distant points;
therefore, they start from distant seed points.
Such seed points are not readily available and must be generated from the
user-supplied seed points in a pre-processing step.
This step is described in more detail in subsection \ref{subsec:bootstrapping}.
Next, the spatial indexing method is modified to allow concurrent insertions of
distant points.
It is divided into two levels, with seed points inserted into the top-level
index and all other points inserted into the bottom level sub-indices.
The top-level can be made read-only and thus thread-safe for the stage in which
the bulk of points is inserted.
The bottom level must use a synchronisation mechanism for the points to be
correctly inserted, but this mechanism can be location-aware.
The global spatial indexing structure is described in more detail in subsection
\ref{subsec:global_indexing}.
The overall pseudocode of the algorithm is presented in
Algorithm~\ref{alg:parallel_algorithm} with the details explained in the
following subsections.

\begin{algorithm}
	\caption{The proposed parallel algorithm.}
	\label{alg:parallel_algorithm}
	\smaller
	\SetAlgoLined
	\SetKwInOut{Input}{input}
	\SetKwInOut{Output}{output}
	\SetKwFor{pfor}{parallel forall}{do}{end}
	\SetKwFor{With}{with}{do}{end}
	\SetKw{Global}{global}
	\SetKwBlock{Begin}{begin}{end}

	\Input{Domain $\Omega$}
	\Input{Nodal spacing function $h$ defined on $\Omega$}
	\Input{A list of initial seed points $X_\text{input}$}
	\Input{Number of seed points to generate, $n_s$}
	\Input{Estimated number of points to cover the whole $\Omega$: $n_p$}
	\Output{A list of points $X$}

	\Begin{
		\Global{$X$} $\gets$ \{\} \tcp*{Location-aware indexed list of points.}

		\tcp{Partition the domain (bootstrapping step).}
		$d \gets $dimensionality of $\Omega$ \;
		$a \gets (n_p / n_s)^{1/d}$ \;
		$h_\text{bootstrap} \gets a\cdot h$ \;
		$\text{s}_b \gets$ new Sequential algorithm \;
		$X_\text{seed} \gets \text{s}_b(\Omega$, $h_\text{bootstrap}$, $X_{\text{input}}$) \;
		topLevelIndex($X$) $\gets X_\text{seed}$

		\{${X_\text{seed,1}, ... X_{\text{seed},n_p}}$\} $\gets$ partition($X_\text{seed}$, $n_p$) \tcp*{divide equally into $n_p$ partitions}

		\tcp{Parallel fill}
		\pfor{$i_p \in [1, n_p]$}
		{
			start $\text{thread}_i$ \;
			\With{$\text{thread}_i$}
			{
				\tcp{Run sequential fill on predefined structures.}
				s $\gets$ new Sequential algorithm \;
				point s.spatialIndexing to $X$ \;
				redirect s.output to $X$ \;
				sequential($\Omega$, $h$, $X_{\text{seed},i}$)
			}
		}
	}
	\Return{$X$}
\end{algorithm}

\subsection{Bootstrapping}
\label{subsec:bootstrapping}

When designing algorithms optimised for efficiency, sometimes additional
requirements and conditions on the input data are imposed, relative to the
simpler algorithm's requirements and conditions.
If those requirements can be relaxed by adding a pre-processing step to
condition the input data without external help, we can call this step
bootstrapping.
In the case of the presented parallel algorithm, the input seed points can be
insufficient in number (one per thread is required) and in their spacing (they
should be as distant as possible to maximise concurrency).
Therefore pre-processing is aimed at generating a better set of seed points.
This step's goals are very similar to the goals of the fill algorithm itself
but with a much lower number of points to place.
So, how could the fill algorithm be modified to bootstrap itself?

The main idea for placing several seed points is simple.
For pre-processing, the algorithm should be executed exactly as for the main
processing, using only a different spacing function $h_\text{bootstrap}$.
To reduce the number of placed points down to the target number of seed points
and to keep relative point density distribution unmodified, the spacing
function should be linearly amplified: $h_\text{bootstrap}=a\cdot h$.
The amplification factor $a$ is used to linearly reduce the number of placed
points per unit of $d$-dimensional volume.
It can therefore be written as
\begin{equation}
\label{eq:bootstrap_estimate_a}
a = (n_p / n_s)^{1/d}.
\end{equation}
Note that the total number of placed points is usually not known exactly and
can only be estimated
and that the amplified spacing function cannot capture gradients in point
density on the same level of detail as the original can.
These two factors will cause the algorithm to \q{miss} the target number of
seed points in most cases, but the number of placed points should still be
close to the required number.

\begin{figure}
	\centering
	\includegraphics[domainHalfFig]{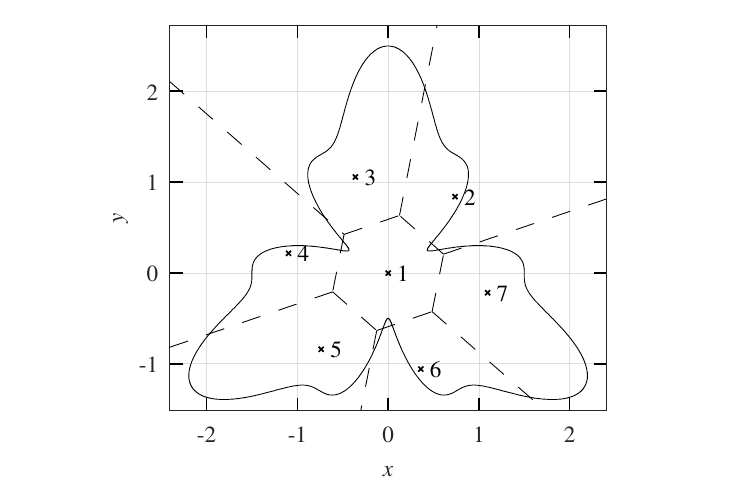}
	\caption{
		The clover domain divided into cells by the application of bootstraping.
		Cell borders are plotted with a dashed line.
		Centre points of cells, marked by $\times$, are seed points.
	}
	\label{fig:cells_illustration}
\end{figure}

The generated seed points can be used to partition the domain $\Omega$ into
sub-domains $\Omega_i$ we call cells, because of their similarity to the
Voronoi cells.
A cell $\Omega_i$, defined by a seed point $p_{i}$, is the area containing
those points from the domain that are closer to $p_i$ than to any other seed
point:
\begin{equation}
\Omega_i = \left\{
x; \;
\|x - p_i\|  = \min_{\substack{j = 1, \ldots, n_s \\ j \neq i}} \|x-p_j\|
\right\}.
\end{equation}
The division into cells can then be used for location-aware synchronisation of
the global spatial indexing.
An example of division is shown in Figure~\ref{fig:cells_illustration}.
Note that even though the goal was to generate 4 seed points for 4 threads on
the given example, the bootstrapping resulted in 7 seed points.
It is trivial for the algorithm to use more seed points than threads, so such a
result is perfectly acceptable.

\subsection{Global spatial indexing}
\label{subsec:global_indexing}

\begin{figure}[ht]
	\includegraphics[fullFig]{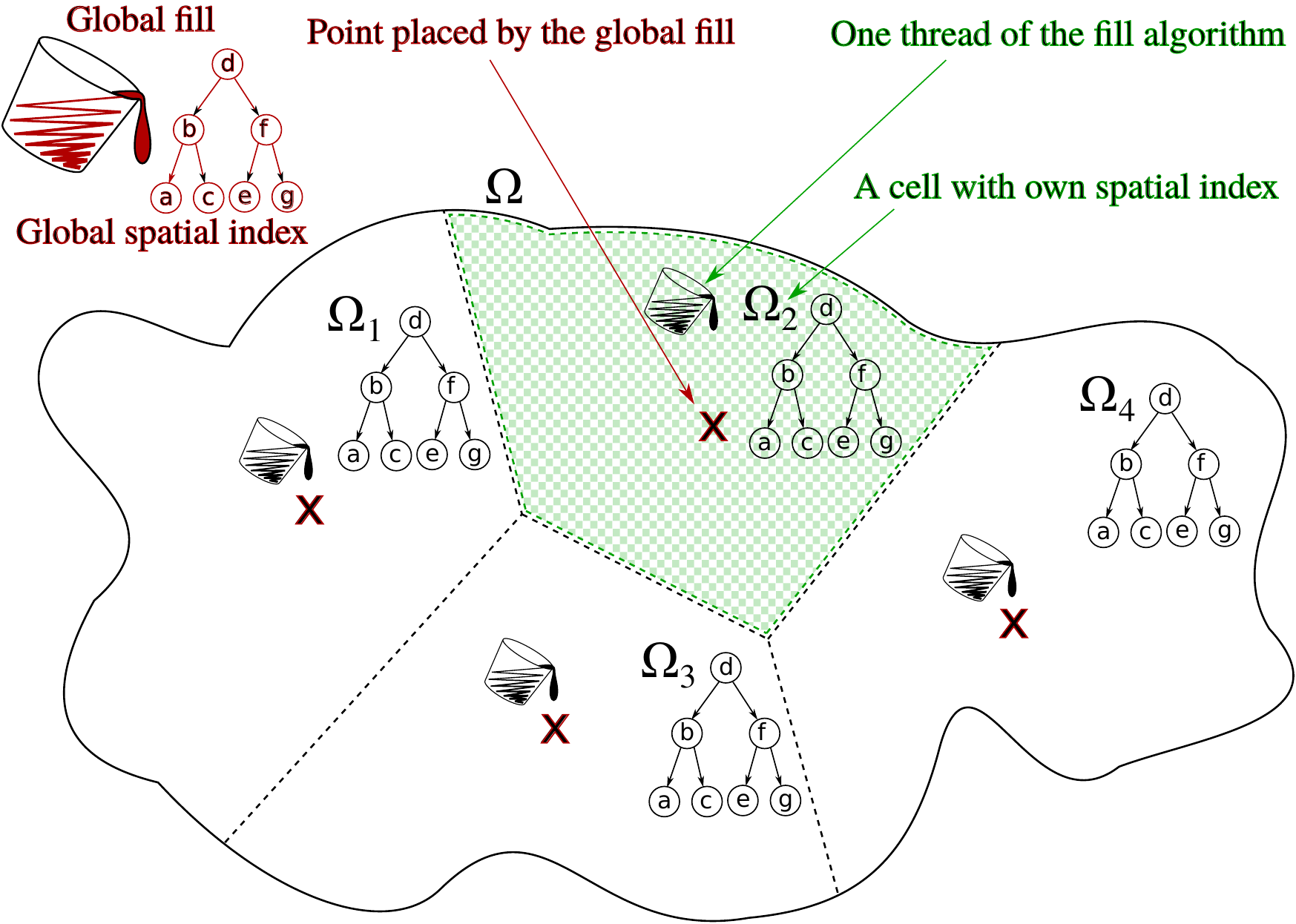}
	\caption{
		A fictional domain divided into cells. Individual fill threads are
		represented by buckets and spatial search indexes as binary trees.
	}
	\label{fig:algorithm_illustration}
\end{figure}

As explained before, spatial indexing is divided into two levels.
The two levels differ by the type of points they hold; the top-level is for
seed points, while the bottom level is for all other points.
The top-level index $S_{top}$ is generated during the bootstrapping phase,
using a sequential algorithm.
It is then made read-only -- the points it holds define the cells, which form
the base of the bottom level and may not be modified during the algorithm's
execution.
The bottom level is divided into as many sub-indices as there are seed points.
Each sub-index $S_i$ is used to index the space covered by its cell $\Omega_i$.
Sub-indices can also be thought of as branches if spatial indexing is
performed by a $k$-d tree.
The illustration of the two-level spatial index is shown in
Figure~\ref{fig:algorithm_illustration}.
Note that the illustrated modifications can be applied regardless of the
spatial indexing method used, which was $k$-d tree in our case.

Using the global spatial index, the procedure for placing a candidate point $c$
is transformed as follows.
First the nearest neighbour $p_{nn}$ of $c$ is found among the seed points on
the top-level index.
This can be done concurrently for any number of threads since the index is
read-only. The neighbour
$p_{nn}$ determines the cell $\Omega_{nn}$ and its spatial index $S_{nn}$ in
which the proximity check for the $c$ is then completed.
If successful, $c$ is added to the set of active points and indexed in the same
spatial index, $S_{nn}$.
Access to $S_{nn}$ is synchronised with the use of a readers-writer lock to
protect the index's shared state between a search and insertion.
Readers-writer lock (also called single-writer or multi-reader lock)~\cite{raynal2012concurrent} is a locking mechanism that allows an unlimited number of
concurrent readers but exclusive access for writers.
Concurrent modification to $S_i$ and $S_j$, however is possible when $i \ne j$.
This is the main property that allows the fill algorithm to achieve high
parallel efficiency.

The two-level spatial index can also be used as an additional result of the
fill algorithm, for example to aid in computation of stencils. It can be used
in the same way the traditional spatial index would be, as an immutable
structure that is used for efficient search of $n$ nearest neighbours.
Procedure for searching one or more nearest neighbours in a two-level spatial
index is comparable to that in a single-level spatial index. Although it is
performed in two steps, one for each index level, the total number of operations
is comparable. For example in $k$-d trees, to reach an element in the leaf,
the combined depths of top-level tree and one of the lower-level trees have to
be traversed, which sum up to approximately the total depth of a regular $k$-d
tree index. Furthermore, when used only for searching, the two-level index
can be safely used in parallel without a locking mechanism.

\subsection{Advancing fronts}

A single thread of the parallel algorithm performs an advancing front algorithm
and will place points locally around the seed point.
If more than one seed point is assigned to a thread, then such a thread will be
advancing more than one front, but all the fronts will be localised to their
starting cells in the beginning.
Threads are oblivious to the details of the global spatial indexing and are not
aware of which cell they are working on at any given moment.
Fronts will thus, in time, cross cell borders unimpeded and enter neighbouring
cells.

The efficient use of this method depends on the threads operating on different
cells.
While this is prescribed at the beginning when they operate on different seed
points, it is no longer guaranteed after fronts cross cell borders.
Eventually, parts of fronts will enter other cells and distribute threads'
workload among more than just their starting cells.
Two or more threads will sometimes try to enter critical sections of the same
spatial index simultaneously and will be stopped by locks, introducing wait
times that will lower the overall algorithm efficiency.
It will be difficult for more than a few threads to be locked out and idle at
any given moment in time, though, since their workloads will be distributed
across many cells.

\begin{figure}
	\centering
	\begin{subfigure}{.49\textwidth}
		\includegraphics[domainSubfig]{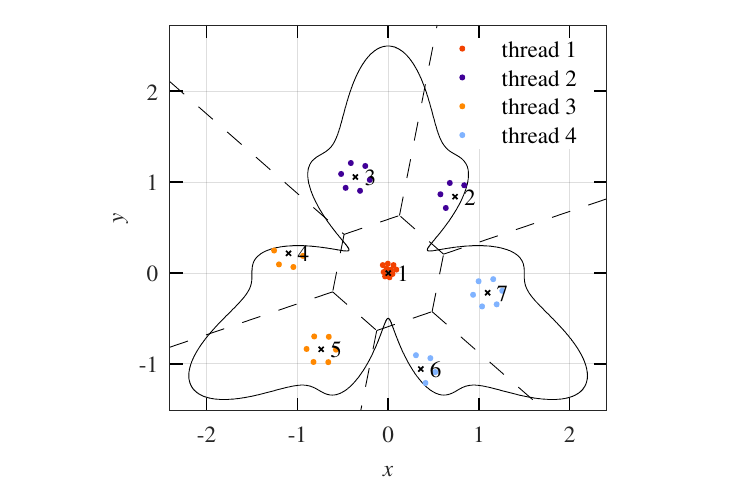}
		\vspace{-10pt}
		\subcaption{}
		\label{subfig:fronts_10_points}
	\end{subfigure}
	\begin{subfigure}{.49\textwidth}
		\includegraphics[domainSubfig]{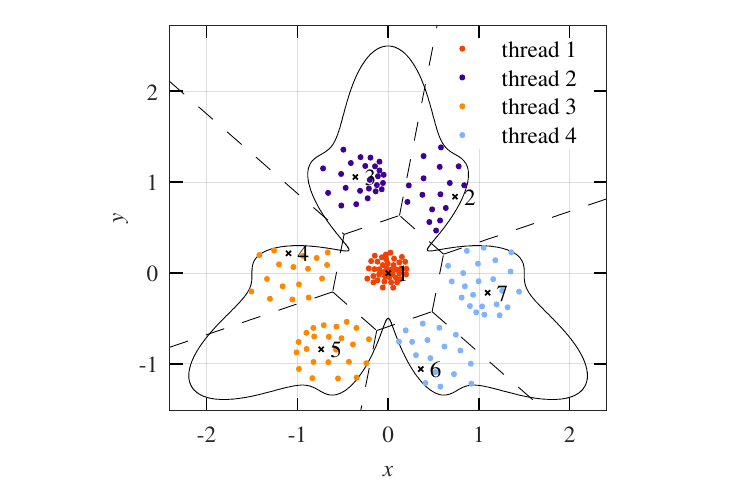}
		\vspace{-10pt}
		\subcaption{}
		\label{subfig:fronts_40_points}
	\end{subfigure}
	\begin{subfigure}{.49\textwidth}
		\includegraphics[domainSubfig]{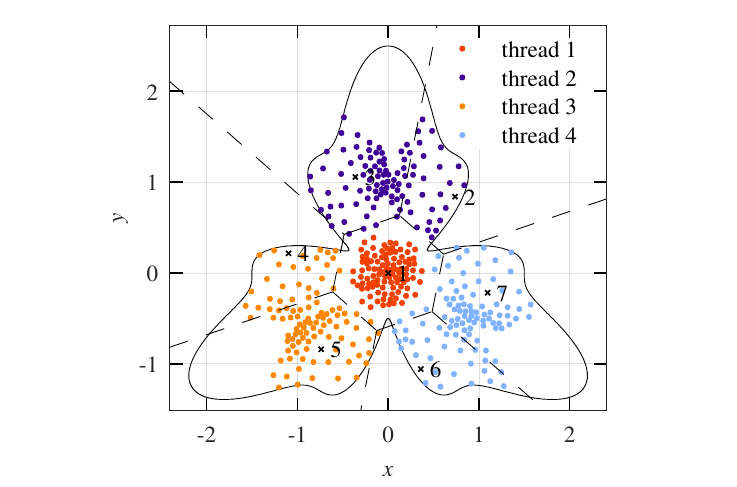}
		\vspace{-10pt}
		\subcaption{}
		\label{subfig:fronts_100_points}
	\end{subfigure}
	\begin{subfigure}{.49\textwidth}
		\includegraphics[domainSubfig]{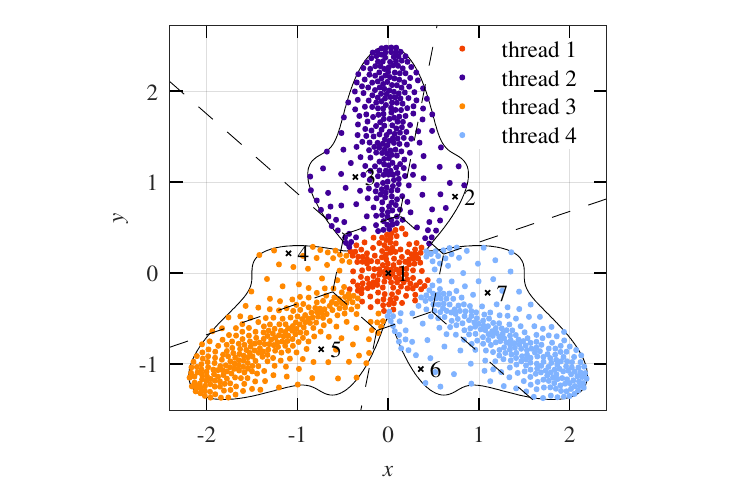}
		\vspace{-10pt}
		\subcaption{}
		\label{subfig:fronts_complete}
	\end{subfigure}

	\caption{
		The clover domain filled with 4 threads and divided into 7 seeds by the
		bootstrapping.
		On all figures, the domain border is plotted with black continuous
		line, cell borders with black dashed line, and seeds by black crosses.
		Four snapshots of the algorithm are depicted: after each thread places
		10, 40, 100 points, and after the algorithm finishes.
	}
	\label{fig:algorihtm_illustration}

\end{figure}

In Figure~\ref{fig:algorihtm_illustration}, an example of algorithm execution
is shown.
Note again that although the cell borders are marked in the figure, these are
not known in advance, and only cell centres are available to the algorithm.
On the presented example, where the number of threads, which is 4, is not a
multiple of the number of seed points, which is 7, seed points are distributed
among the threads unequally.
Three threads start with 2 seed points each, while one starts with a single
seed point.
Thread 1 only gets seed point 1 and starts advancing a single front.
The other three threads initially start with two advancing fronts each, which
can be seen to collide for all three threads in
Figure~\ref{subfig:fronts_100_points}).
The same figure also shows how the fronts advance over the borders of cells.

\subsubsection{Computational complexity}

We define the time complexity relative to the sequential algorithm complexity
$T_\text{sequential}$ from \eqref{eq:complexity_sequential}. The parallel algorithm complexity can be constructed from equation
\begin{equation}
T_{\text{parallel}} = T_{\text{single thread}} + T_{\text{thread}} +
T_{\text{bootstrap}},
\end{equation}
and we explain the additional elements below.
$T_{\text{single thread}}$ is the execution time of the longest running worker
thread.
The worst-case scenario can be conceived where a single thread does all the
work, and therefore this time is a sum of $T_\text{sequential}$ and the
overhead of mutex acquisition.
The latter is bounded by the number of candidate points considered, is constant
per point, and is, therefore, $O(n_{ct} n_p)$.
This is of a lower order than $T_\text{sequential}$, an thus $T_\text{single
	thread}=\Theta(T_\text{sequential})$.
Thread management overhead is specified as $T_\text{thread}$, which depends
linearly on the number of threads $p$ and is constant per thread, thus $O(p)$.
Finally, $T_{\text{bootstrap}}$ is the bootstrap overhead, which equals the
execution of the sequential algorithm for placing the given number of seed
points: $O(n_{ct} n_s \log(n_s))$.
We can further assume a linear relation between $n_s$ and $p$ and therefore
write that $T_{\text{bootstrap}}=O(n_{ct} p \log(p))$.
Combining the elements of the equation gives:
\begin{eqnarray}
T_{\text{parallel}} & = & O(T_\text{sequential} + p + n_{ct} p \log(p))
\nonumber \\
& = & O(n_{ct} n_p \log(n_p) + n_{ct} p \log(p)),
\label{eq:parallel_complexity}
\end{eqnarray}
which is worse than the complexity of sequential algorithm.
This result demonstrates that in the worst-case scenario, the proposed parallel
algorithm could be very inefficient.
An example of such a case is difficult to imagine though.
The domain shape and its nodal spacing function would have to be degenerate to
the point of not being useful for a numerical approach at all.
While it seems improbable that such domains would be created and used by a
human user, they could be created by an automatised method for optimisations or
statistical calculations and should not be entirely disregarded.

To arrive at a more realistic estimate of complexity, we start with an
assumptions of $p \ll n_p$, equally fast processor cores, and $T_\text{single
	thread}=T_\text{sequential}/p$.
The latter is in line with the assumption listed already for the sequential
algorithm, i.e.\ that the values of the spacing function $h$ are at least an
order of magnitude smaller than the sizes of geometric features of $\Omega$.
This pair of assumptions ensures that bootstrapping generates sufficiently
distant seed points for the static load-balancing to work well enough.
Then the time complexity of the parallel algorithm from
\eqref{eq:parallel_complexity} can be rewritten as:
\begin{eqnarray}
T_{\text{parallel}} & = & O(n_{ct} n_p \log(n_p)/p + n_{ct} p \log(p)).
\end{eqnarray}

The spatial complexity becomes bounded by how much space is needed to store
$n_p$ points across a number of spatial indices $S_i$, and by the overhead of
thread management, which is $O(p)$.
In case $k$-d trees are used for spatial indexing, the space complexity is
$O(n_p/p)$ per each of the $p$ spatial indices and the space complexity becomes:
\begin{eqnarray}
S_{\text{parallel}} & = & O(p n_p/p + p) \nonumber \\
& = & O(n_p).
\end{eqnarray}

\section{Experiments}
\label{sec:experiments}

In this section, we experimentally evaluate how well does the presented
parallel algorithm work on multi-core hardware.
We define a set of experiments to evaluate the scalability as a function of a
number of threads and problem size.
A number of threads is used on a range that is available on the experimental
hardware, $p \in [1, 32]$.
Problem size is defined by the total number of points placed $n_p$ and will
range in our experiments from 1000, increasing by a factor of 4, up to 16
million.
Note that our experiments only place approximately the given number of points,
the exact number depending on the random seed, the number of threads, and
positions of user-defined seed points.
Therefore we shall talk about the \q{target} numbers of points, i.e.\ the
number used to calculate the domain spacing function parameters in a way that
should result in approximately that many points placed.
The functions for calculating the spacing function parameters were manually
preliminary fitted to several executions of the algorithm on several problem sizes.
This was done offline and is out of the scope of the article.
Note that the execution times were normalised with the actual number of placed
points to remove the bias originating in a slightly varying number of points
placed by the different runs for calculation of speedups.

\subsection{Experimental setup}
\label{sec:experimental_setup}

A computer based on a AMD Threadripper 2950X, a 16-core 32-thread processor and
32 GB of RAM was used to perform the presented experiments.
The double thread count with respect to the number of cores is the consequence
of the processor's simultaneous multi-threading ability (SMP), which allows a
single physical core to execute two threads concurrently.
The computer was running Ubuntu 20.04 desktop and was kept idle apart from
running the experiments.
As the majority of modern CPUs, the 2950X is capable of changing its clock
frequency on per-core basis in real-time, on the one hand, to ensure maximum
performance, and on the other hand, to limit power usage and keep die
temperature within safe bounds.
To measure the speedup on equally powerful CPU cores, regardless of how many
cores are in use, we have disabled the dynamic frequency boost and set a fixed
frequency of 2.2 GHz for all cores.
Furthermore, we used thread binding in all our experiments, which means that
the number of threads always matches the number of used cores, since each
thread is bound to a single core, and there is at most one thread per core.
Here we also count the virtual cores, i.e.\ the use of SMP results in Linux
seeing the number of cores as twice the number of physical CPU cores.
These virtual cores are numbered from 17 to 32, and we use them only for
experiments with 17 or more threads.
Not using these cores on a lower number of threads improves the use of local
core caches, and thus the performance of the proposed algorithm can nearly
fully utilise any core with a single thread.

All the presented algorithms were implemented in C++, using its standard
library for multi-threading (\verb|std::thread|).
Threads were created with C++ lambda expressions, which deferred the call to a
member function on a pre-constructed object instance.
The end of thread execution was detected with the \verb|std::thread::join|
function, which suspends the waiting thread until the target thread completes.
Mutexes of class \verb|std::shared_mutex| were used, which implement two levels
of access protection - shared and unique, which makes it possible to protect
critical sections of code with readers-writer locks.
The spatial indexing was performed using a mature and heavily optimised library
\q{nanoflann}, which supports both statically and dynamically constructed $k$-d
trees \cite{blanco2014nanoflann}.
For the top-level index, a statically built tree was used, while for the bottom
level sub-indices and for the sequential algorithm, the dynamically built trees
were used.

\subsection{Speedup}

\begin{figure}[ht]
	\centering
	\includegraphics[fullFig]{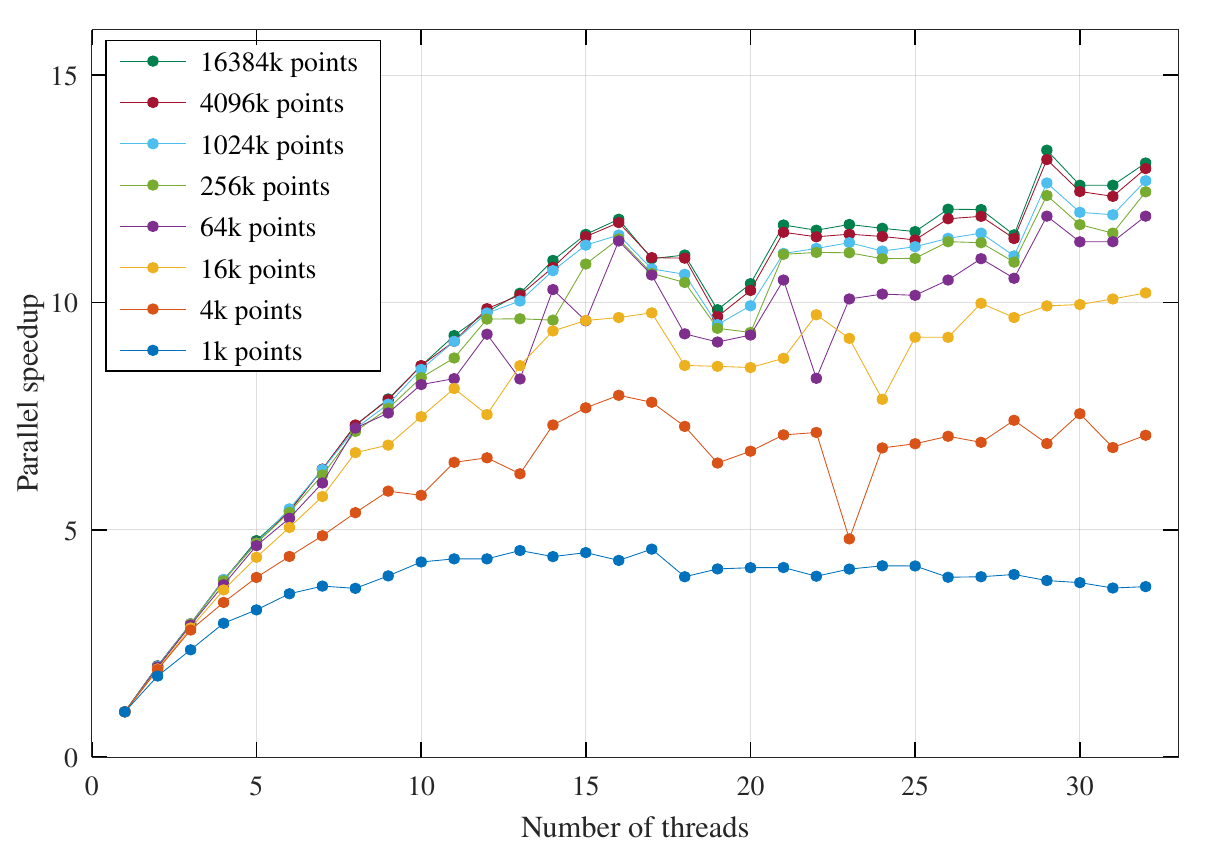}
	\caption{The speedup of the cell-based parallel algorithm as a function of
	the number of threads for several problem size levels, defined by the
	number of placed points. All the results were obtained on the 2-D clover
	domain.}
	\label{figure:cellPerformance}
\end{figure}

Speedups from all the performed experiments, computed relative to the
sequential algorithm, are shown in Figure~\ref{figure:cellPerformance}.
There are several main observations possible from this figure.
Foremost, the speedups are significant, especially for the large problem sizes.
Moreover, they are mostly increasing up to 16 threads, which is the number of
cores on the computer used, and nearly linear for the low number of threads.
The use of additional threads, available due to the processor's support for
SMP, ranges in effect from slightly hindering performance to slightly boosting
performance.
The likely cause of this is that the algorithm's workload is not diverse enough
for SMP to be able to run two threads on the same core efficiently.
Running the algorithm on 17 threads, for example, thus loads physical core 1
with two threads and all the others with only one, and thus makes the two
threads on core 1 run appreciably slower than the rest.
The effect of this is detrimental since the algorithm assumes equally fast
execution on all threads, which no longer holds in the described situations.
A dynamic load balancing could be implemented to tackle this issue, but this is
beyond the scope of the presented algorithm.
In further tests and discussions, we focus mostly on the performance when all
physical cores are loaded, that is, at 16 threads

It is also very apparent that placing even a very low number of points can be
sped up using multi-threading, and the only instance of the problem which does
not scale well to at least 16 threads is 1000 points.
More substantial speedups are only achievable when several ten thousand points
are to be placed, and the best-achieved result in the shown experiments was a
speedup of 12 at 16 threads.
To get a better feeling for the problem sizes, it takes the sequential
algorithm 12~ms to place 1000 points and 230~s to place 16 million points.

\subsection{Domain shape and spacing}
\label{sec:domain_shape_and_spacing}

\begin{figure}[ht]
	\centering
	\subframe{
		\begin{subfigure}{.37\textwidth}
			\includegraphics[height=4.7cm, trim=15mm 1mm 16mm
			2mm,clip]{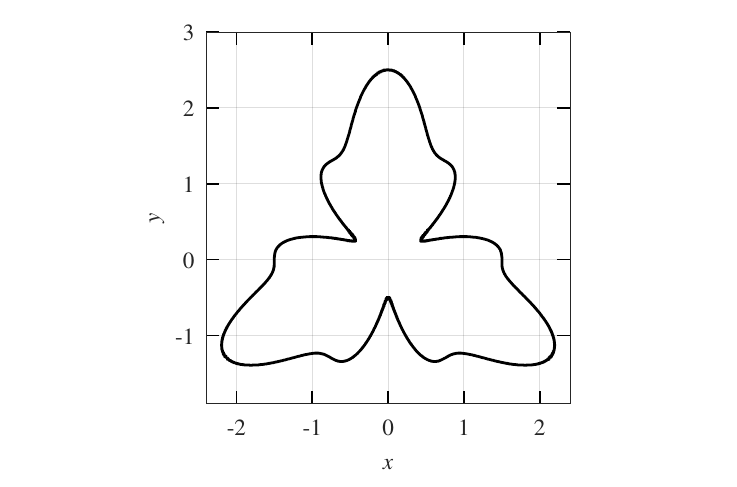}
			\subcaption{2-D clover}
		\end{subfigure}
	}
	\subframe{
		\begin{subfigure}{.48\textwidth}
			\includegraphics[height=4.7cm, trim=8mm 1mm 10mm
			2mm,clip]{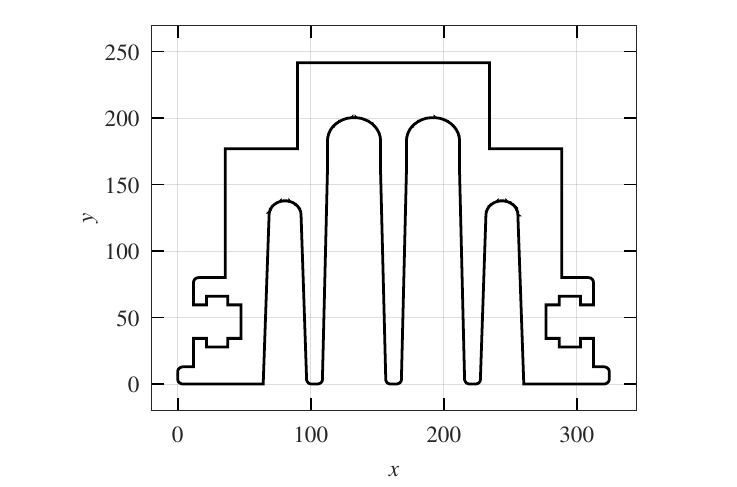}
			\subcaption{2-D heatsink}
		\end{subfigure}
	}
	\subframe{
		\begin{subfigure}{.31\textwidth}
			\includegraphics[height=4.7cm, trim=19mm 1mm 19mm
			2mm,clip]{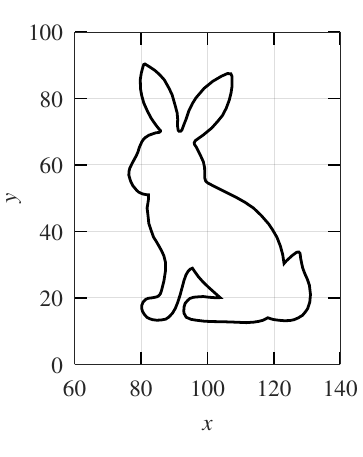}
			\subcaption{2-D bunny}
		\end{subfigure}
	}
	\subframe{
		\begin{subfigure}{.53\textwidth}
			\includegraphics[height=4.7cm, trim=5mm 1mm 7mm 2mm,clip]{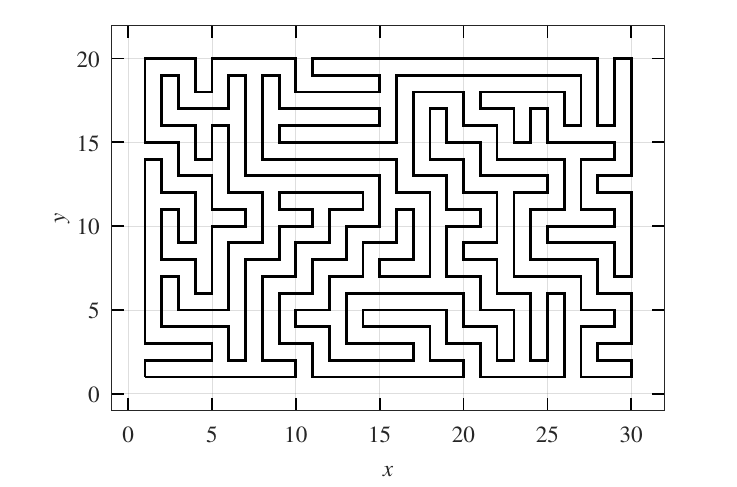}
			\subcaption{2-D maze}
		\end{subfigure}
	}

	\caption{2-dimensional test domains.}
	\label{fig:test-domains-2d}
\end{figure}

\begin{figure}[ht]
	\centering
	\subframe{
		\begin{subfigure}{.41\textwidth}
			\includegraphics[height=4.4cm]{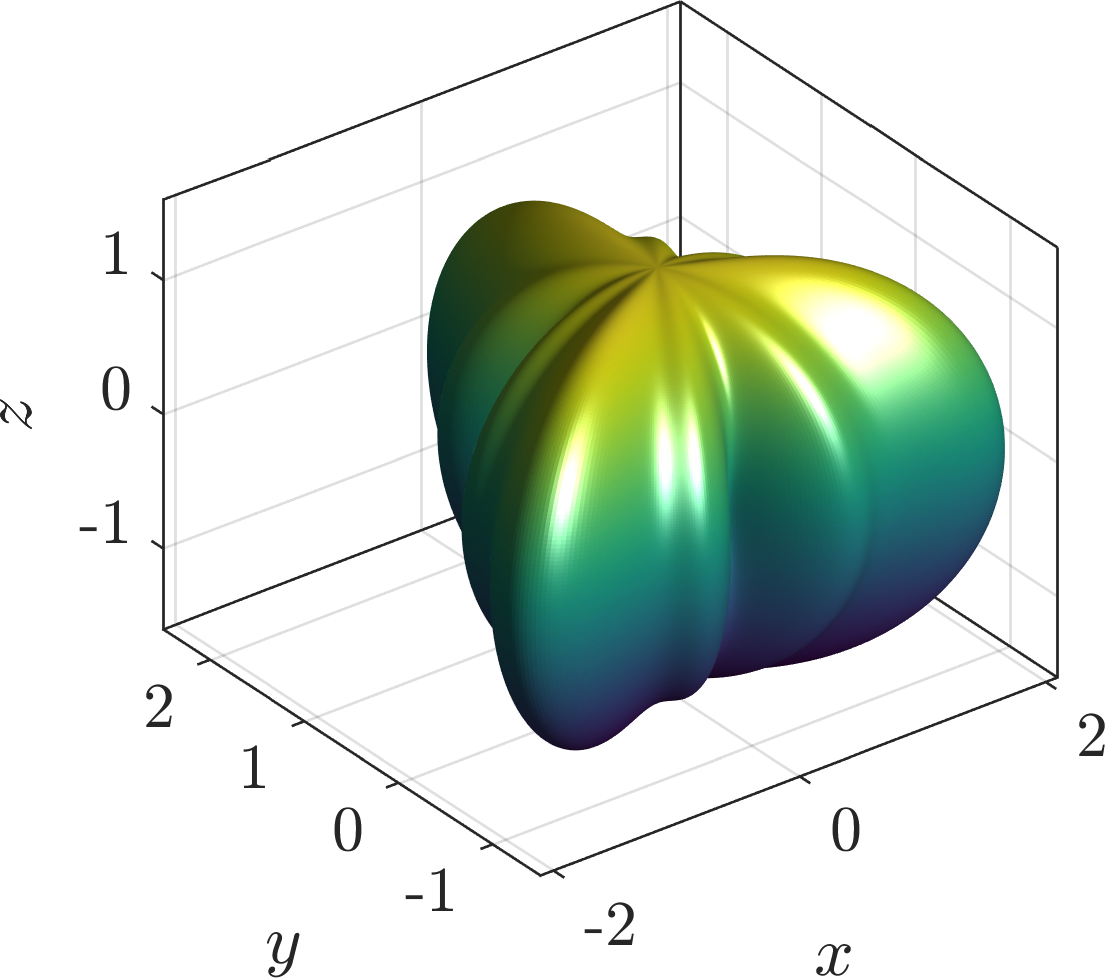}
			\subcaption{3-D clover}
		\end{subfigure}
	}
	\subframe{
		\begin{subfigure}{.49\textwidth}
			\includegraphics[height=4.4cm]{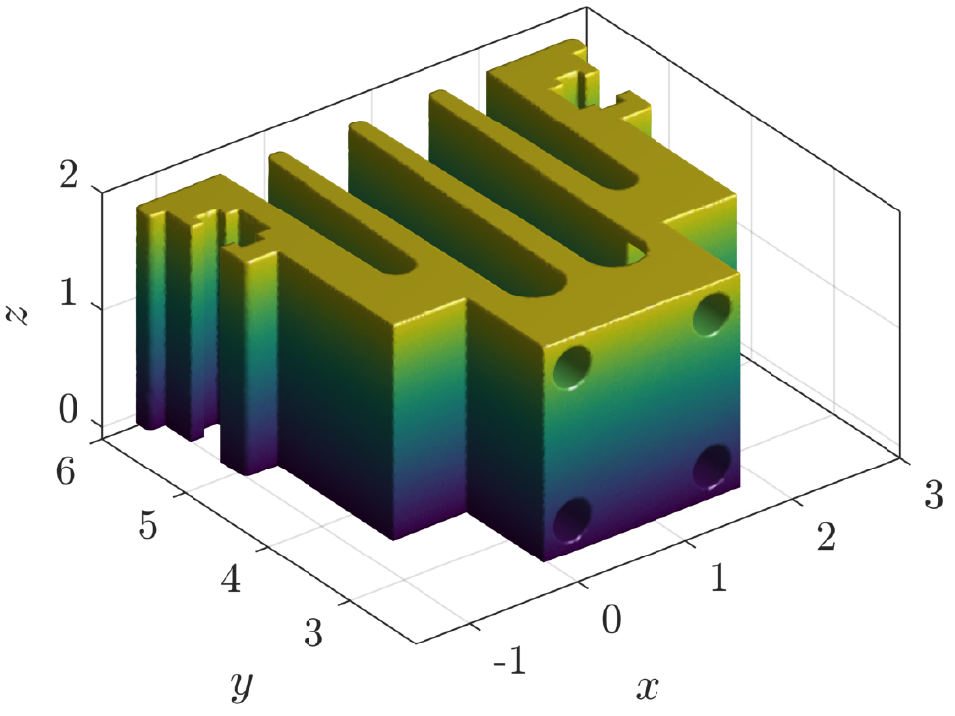}
			\subcaption{3-D heatsink}
		\end{subfigure}
	}
	\subframe{
		\begin{subfigure}{.38\textwidth}
			\includegraphics[height=4.6cm]{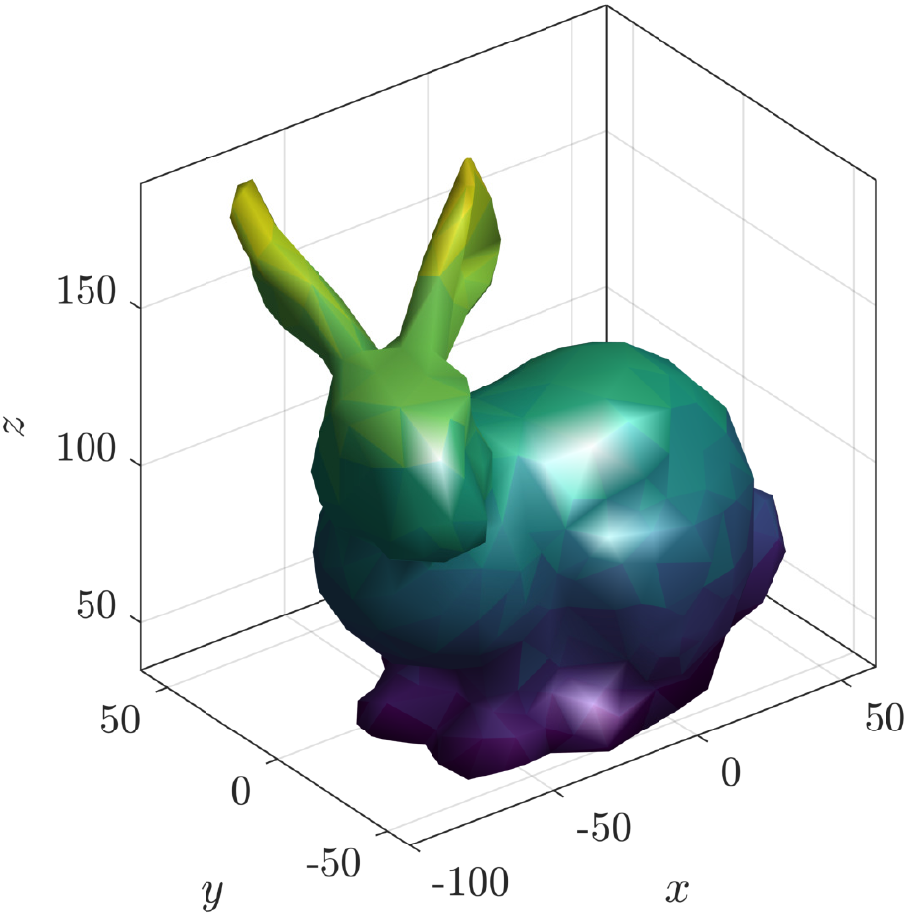}
			\subcaption{3-D bunny}
		\end{subfigure}
	}
	\subframe{
		\begin{subfigure}{.40\textwidth}
			\includegraphics[height=4.6cm]{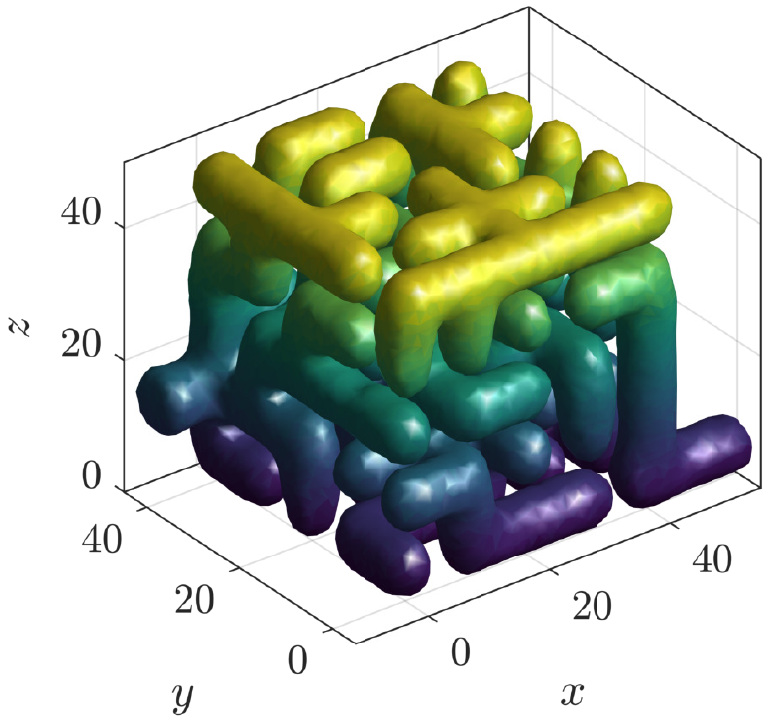}
			\subcaption{3-D maze}
		\end{subfigure}
	}
	\caption{3-dimensional test domains.}
	\label{fig:test-domains-3d}
\end{figure}

Eight domains are used to demonstrate the robustness of the algorithm.
These domains can be classified into four pairs: \q{heatsink}, \q{clover},
\q{bunny} and \q{maze}, each in 2-D and 3-D version.
The 2-D domains are shown in Figure~\ref{fig:test-domains-2d} and 3-D domains
in Figure~\ref{fig:test-domains-3d}.
The 2-D clover domain was already used in visualisations of the algorithms and
the 3-D heatsink is the domain used in Section~\ref{sec:solving_pde}.
The 3-D heatsink is the only domain which is not homeomorphic to a sphere.

Both heatsink domains are used in conjunction with the uniform point spacing,
which seems optimal for the demonstrated simulation of heat transfer.
Since non-uniform nodal distributions are often used in real-life domains and
are also often the source of performance loss in other node placement
algorithms, we have devised non-uniform spacing functions for the other 6
domains.

The boundaries of both 2-D and 3-D versions of clover
are defined in closed form, with 2-D defined in~\eqref{eq:clover} and 3-D
given as a surface in spherical coordinates as
\begin{equation}
r_{\text{clover-3D}}(\phi, \theta) =
\frac32 - \cos(3 (\phi - \pi/6))^3 (\pi - \theta)^2 \theta^2/8, \quad
\phi \in [0, 2\pi),\; \theta \in [0, \pi).
\end{equation}
Both versions of clover are also star-shaped with respect to the origin,
which allows us to easily define the characteristic function by just checking
the inclusion radius.

The rest of the domains are polytopes. The 3-D bunny is a watertight version of
the standard Stanford bunny model~\cite{bunny}. The 2-D version was drawn
manually in InkScape, from where the coordinates were also exported. The bunny
domains are used to showcase the behaviour of the algorithm on realistic
2-D and 3-D objects. Both of the maze domains are meant to explore the behaviour
of the algorithm on more complex structures. The 2-D version was generated using
a quick recursive-backtracking maze generation algorithm, and the 3-D version
was generated using~\cite{maze3d} with some post-processing in Meshlab.
The 2-D heatsink domain is the projection of 3-D heatsink on the $xy$-plane.

Spacing functions for the presented domains are defined in table~\ref{tab:spacing}.

\begin{table}
  \caption{Different spacing functions for the presented domains.}
  \label{tab:spacing}
\begin{eqnarray}
\textrm{2-D}	& & h_t(x, y) = \nonumber \\
\hline \nonumber \\[-1pt]
\textrm{heatsink} 	& & \hm \\
\textrm{clover}	& &  \hm \left( 1+4 \cos^2(3\arg(x, y))\tanh(\sqrt{x^2+y^2})
\right) \\
\textrm{bunny}	& & \hm \left( (1+y/100)^{1.5} \right) \\
\textrm{maze} 	& & \hm \left( (1+y/20)^{1.5} \right)	\\[10pt]
\textrm{3-D}	& & h_t(x, y, z) = \nonumber \\
\hline \nonumber \\[-1pt]
\textrm{heatsink} 	& & \hm \\
\textrm{clover}	& & \hm \left( 0.5+\cos^2(3\arg(x,
y)+\pi/3)\tanh((2-z)\sqrt{x^2+y^2+z^2}) \right)	\\
\textrm{bunny}	& & \hm \left( 1+(4*(180-z)/180) \right)	\\
\textrm{maze}	& & \hm \left( 4+\sin(x \pi / 5) \right)
\end{eqnarray}
\end{table}
Equation for the 2-D clover spacing function is derived
from~\eqref{eq:clover-den}, by requiring ${h_{max}}/{h_{min}}=5$, and then
replacing $h_{min}$ and $h_{max}$ with a single scaling parameter $\hm$.
The latter is used to adjust for the total number of placed points, as required
by the experiments.
Similarly all the other spacing functions have a fixed maximal to minimal value
ratio and use a single scaling parameter.
Spacing function for both heatsink domains comprises only a scaling factor,
since these two domains employ a uniform spacing.

\begin{figure}[ht]
	\centering
	\subframe{
		\begin{subfigure}{0.41\textwidth}
			\includegraphics[domainSubfig]{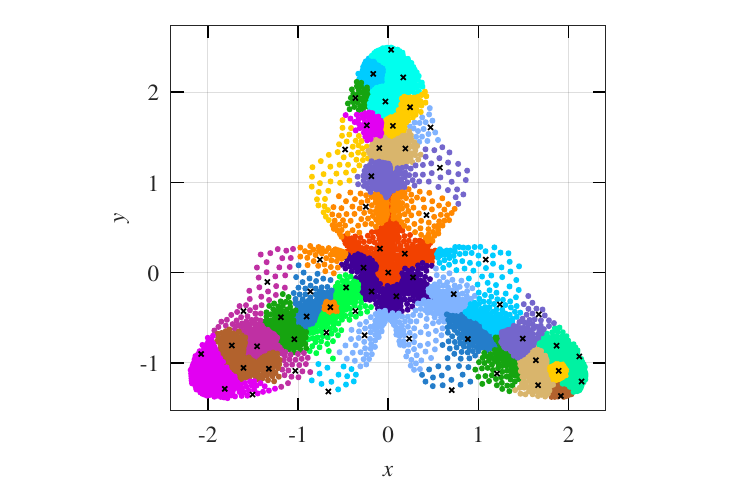}
			\subcaption{2-D clover}
			\label{subfig:filled_domain_clover_2d}
		\end{subfigure}
	}
	\subframe{
		\begin{subfigure}{0.50\textwidth}
			\includegraphics[trim=6mm 0mm 9mm
			2mm,clip,width=\textwidth]{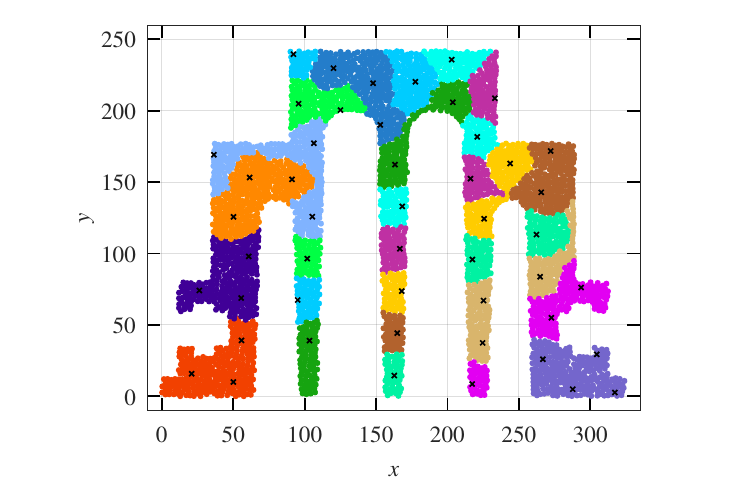}
			\subcaption{2-D heatsink}
			\label{subfig:filled_domain_heatsink_2d}
		\end{subfigure}
	}\\
	\subframe{
		\begin{subfigure}{0.32\textwidth}
			\includegraphics[trim=20mm 0mm 21mm
			2mm,clip,width=\textwidth]{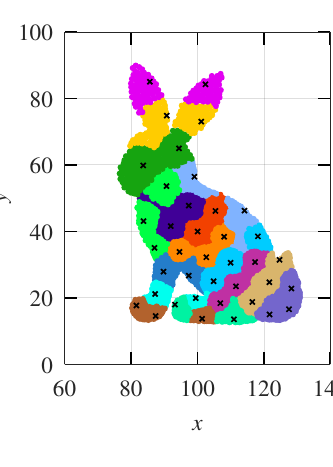}
			\subcaption{2-D bunny}
			\label{subfig:filled_domain_bunny_2d}
		\end{subfigure}
	}
	\subframe{
		\begin{subfigure}{0.60\textwidth}
			\includegraphics[trim=5mm 0mm 6mm
			2mm,clip,width=\textwidth]{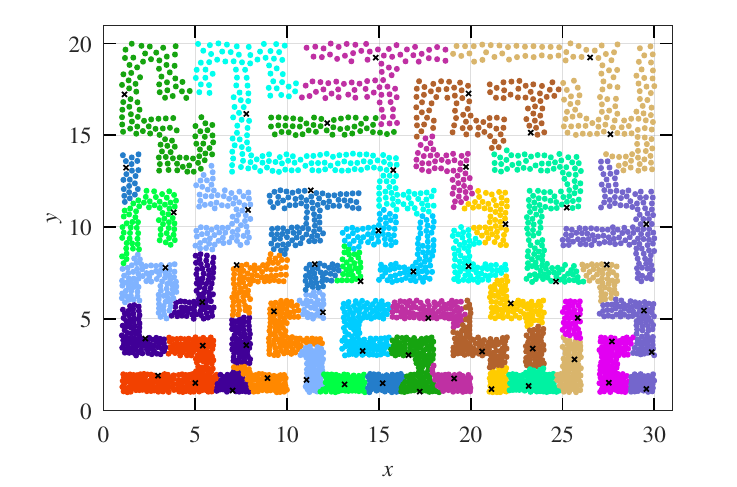}
			\subcaption{2-D maze}
			\label{subfig:filled_domain_maze_2d}
		\end{subfigure}
	}
	\caption{The 2-dimensional domains, filled with cell-based algorithm
	running on 16 threads.
		Points placed by different threads are coloured by different colours.
		While the variable spacing is clearly seen in the clover and maze
		domains, it is also indicated from the variable size of cells in the
		bunny domain, where spacing increases from bunny's feet and tail
		towards its head and ears.}
	\label{figure:otherDomainsInParallel2D}
\end{figure}

\begin{figure}[ht]
	\centering
	\subframe{
		\begin{subfigure}{.44\textwidth}
			\includegraphics[domainHalfFig3d]{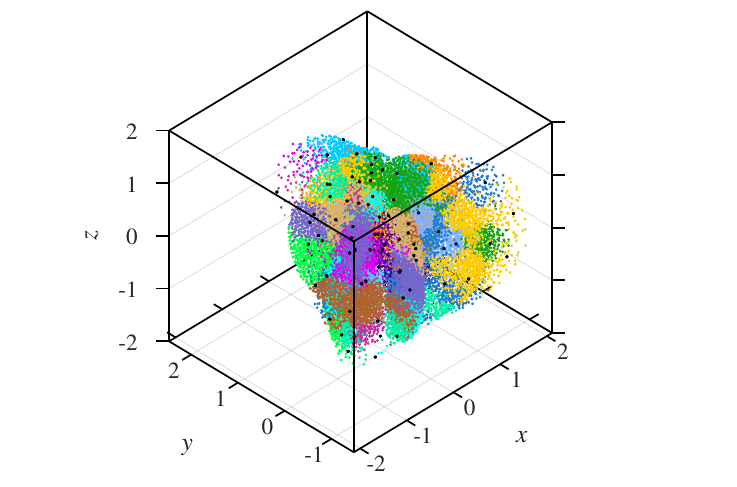}
			\subcaption{3-D clover}
			\label{subfig:filled_domain_clover_3d}
		\end{subfigure}
	}
	\subframe{
		\begin{subfigure}{.44\textwidth}
			\includegraphics[domainHalfFig3d]{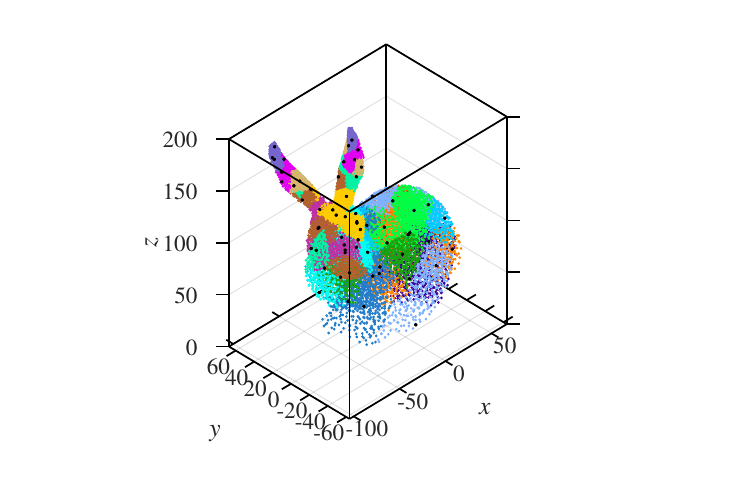}
			\subcaption{3-D bunny}
			\label{subfig:filled_domain_bunny_3d}
		\end{subfigure}
	}
	\subframe{
		\begin{subfigure}{.44\textwidth}
			\includegraphics[domainHalfFig3d]{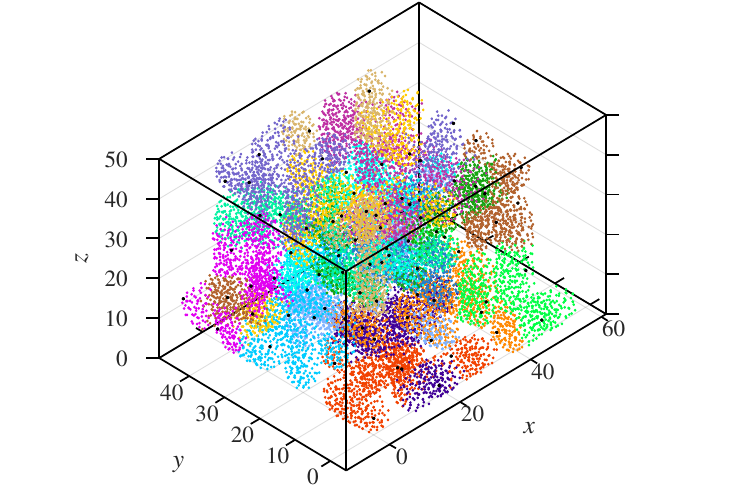}
			\subcaption{3-D maze}
			\label{subfig:filled_domain_maze_3d}
		\end{subfigure}
	}
	\subframe{
		\begin{subfigure}{.44\textwidth}
			\includegraphics[domainHalfFig3d]{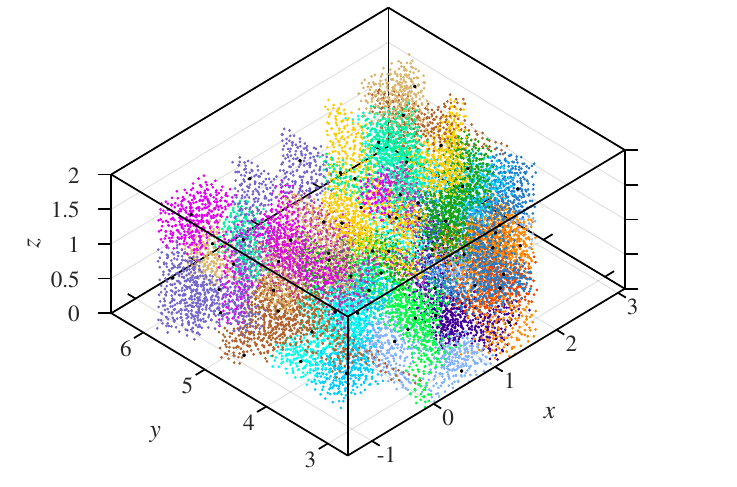}
			\subcaption{3-D heatsink}
			\label{subfig:filled_domain_heatsink_3d}
		\end{subfigure}
	}
	\caption{The 3-dimensional domains, filled with a cell-based algorithm
	running on 16 threads.
		Points placed by different threads are coloured by different colours.}
	\label{figure:otherDomainsInParallel3D}
\end{figure}

We plot one resulting set of scattered nodes for each domain, with colours of
the points corresponding to the threads that placed them.
The 2-dimensional domains were filled with a target $n_p=4000$ points and are
presented in Figure~\ref{figure:otherDomainsInParallel2D}.
The 3-dimensional domains were filled with a target $n_p=16000$ points and are
presented in Figure~\ref{figure:otherDomainsInParallel3D}.
The parameters common to all sub-figures are $n_c=12$, $p=16$, $n_s=32$, and
the use of bootstrapping to generate seed points from a single internal seed
point.
The visualisations of the results show no visible deficiencies, all the domains
are filled with points as well as can be expected.

\begin{figure}[ht]
	\centering
	\includegraphics[fullFig]{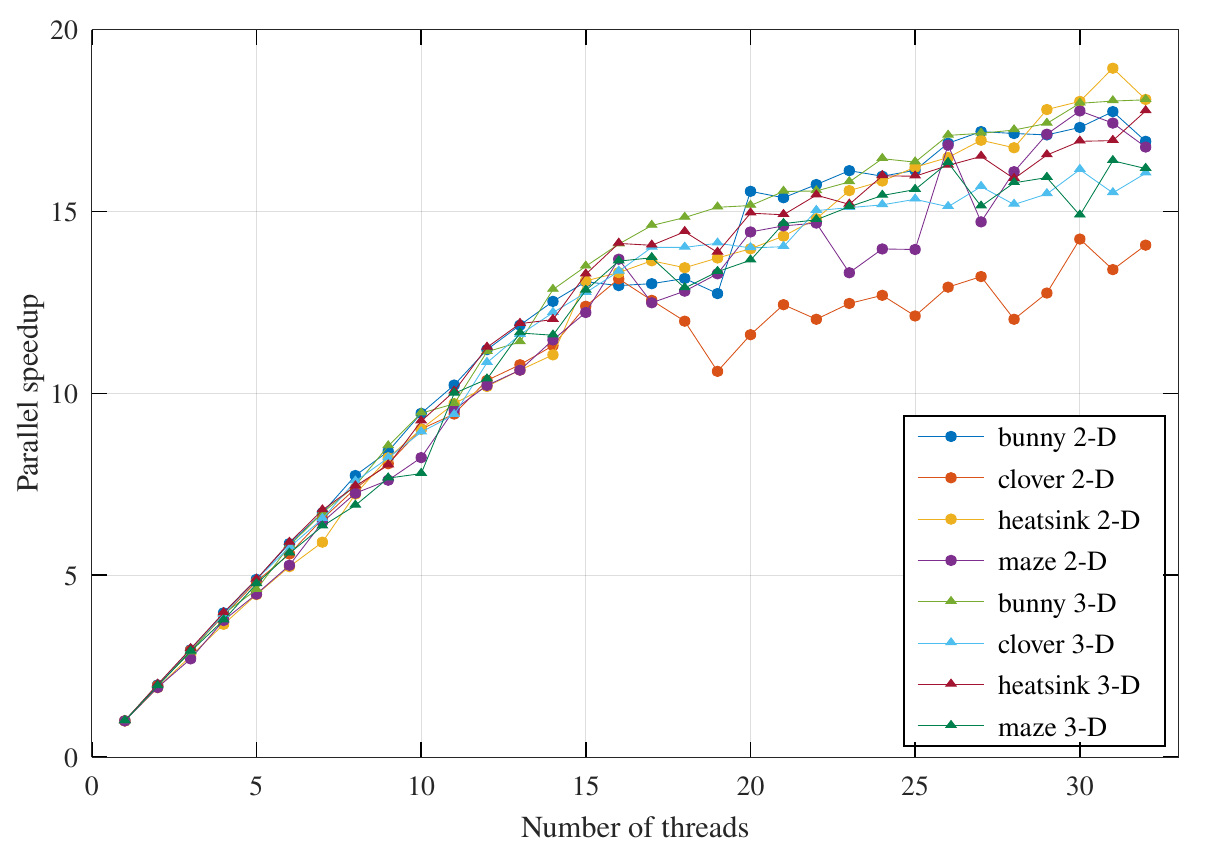}

	\caption{The speedup on various domain types.}
	\label{figure:otherDomainsSpeedup}
\end{figure}

In addition to the visual examination of the results, speedups were calculated
for the parallel algorithm on all the presented domains.
In Figure~\ref{figure:otherDomainsSpeedup}, speedups are shown with fixed
$n_p=4096000$ and varying $p$.
The results reveal that there is very little difference between the 2-D and 3-D
domains in terms of scalability.
One pattern that emerges is that 2-D clover domain performs the worst and even
that is only noticeable on 17 or more threads.

Although the complexity of the domain border, which dictates the time required
to calculate $\Omega$, should also influence the performance, this cannot be
demonstrated on the presented domains.
E.g., the performance does vary a lot on the 2-D maze with its arguably most
complex border shape due to the lack of dynamic load-balancing, but is
generally distinctly better than the performance on the much simpler 2-D clover
shape.

The overall result of the experiments presented in this section is that while
the performance does depend on the domain shape and spacing function, it varies
very little if only the physical cores of the computer are used.
For better exploitation of the SMP, however, a dynamic load-balancing is likely
required.

\subsection{Quality of point placement}

The presented algorithm's end goal is to use the generated points as nodes of a
meshless numerical approach.
The parallel version of the algorithm does not produce exactly the same node
distributions as the sequential version. This may cause some concerns about the
quality of the node distribution produced by the parallel version. However,
the sequential version, run with the set of points obtained during the parallel
algorithm's bootstrapping phase as the seed point set, will produce nearly the
same point set
as the parallel algorithm. The sequential algorithm produces good quality point
distributions, provided that the given set of seed points is of sufficient
quality, which is true in our case, as the
sequential algorithm itself generated the seeds. Nonetheless, we compare some of
the basic quality
measures of node sets for numerical approaches, as used
in~\cite{SlakKosec2019NodeGen}
and~\cite{van2019fast}.

A basic measure of node quality is to check how well the distances to
neighbouring nodes match the prescribed nodal spacing function $h$.
We compute the distances $\{d_{i, j}\}_{i=1,j=1}^{N, k}$ for each node $p_i$ to
its $k$ nearest neighbours (excluding $p_i$ itself), and define the normalized
distances $d_{i,j}' = d_{i,j} / h(p_i)$. A histogram of $d_{i,j}'$ is then
plotted, with the expectation that most values are located around $1$.

\begin{figure}[ht]
	\centering
	\includegraphics[height=4.9cm]{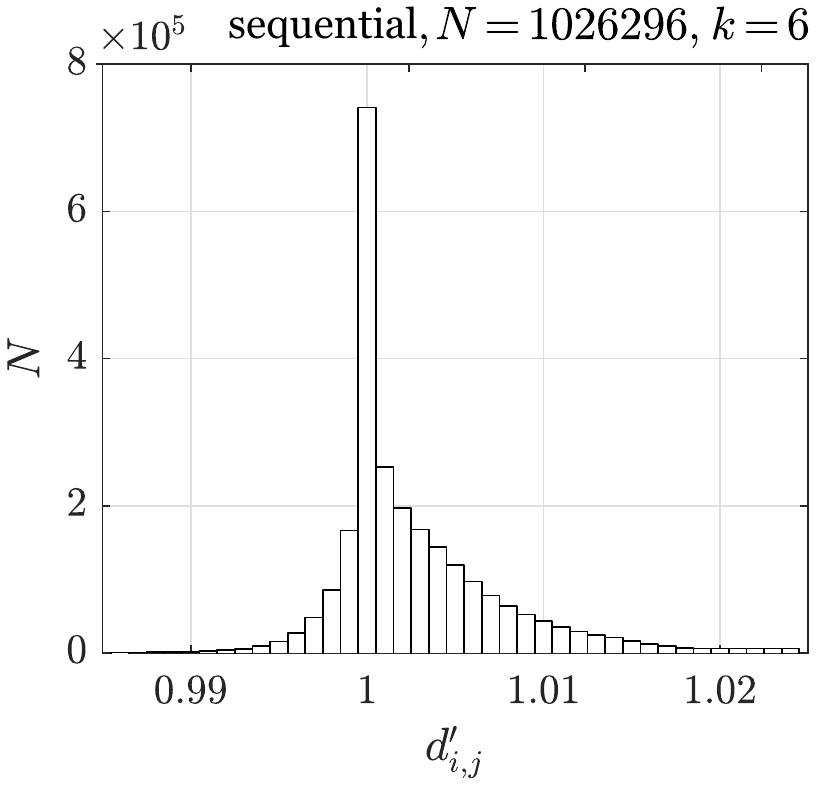}
	\includegraphics[height=4.9cm]{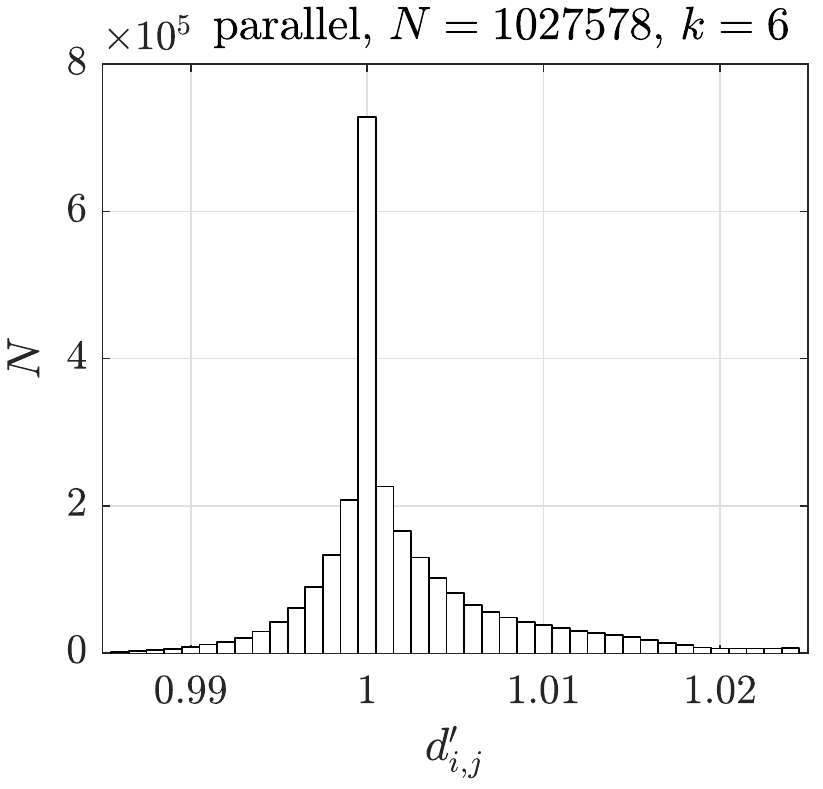}
	\caption{Histograms of normalized distances for sequential and parallel versions
		of the algorithm.}
	\label{fig:node-hist}
\end{figure}

Such histograms are shown in figure~\ref{fig:node-hist} for the 2-D clover
shape~\eqref{eq:clover} with the nodal spacing function~\eqref{eq:clover-den},
with $\hmin = 0.0016$ and $\hmax = 0.0078$. There are some differences between
the histograms of the sequential and parallel version, but they are an order of
magnitude smaller than the actual counts, and both histograms represent
acceptable
node distributions.

Additionally, we can calculate the average distance to nearest neighbours
$\bar{d}_i = \frac{1}{k} \sum_{j=1}^{k} d_{i,j}$, as well as the maximum and
minimum distances to neighbours for each point:
\begin{equation}
d_i^{\text{min}} = \min_{j=1, \dots, k} d_{i,j}, \qquad
d_i^{\text{max}} = \max_{j=1, \dots, k} d_{i,j}.
\end{equation}
All these distances can be normalized by $h(p_i)$, to obtain $\bar{d}'$,
$(d_i^{\text{min}})'$, etc. Some statistics of $d_{i,j}$ are recorded in
table~\ref{tab:node-stat}. Both algorithms perform well, with the parallel
version showing slightly more dispersed distances to nearest neighbours.

\begin{table}[ht]
	\renewcommand{\arraystretch}{1.2}
	\centering
	\caption{Numerical quantities related to local regularity.}
	\label{tab:node-stat}
	\begin{tabular}{c|c|c|c}
		algorithm & $\operatorname{mean}\bar{d}'_i$ &
		$\operatorname{std}\bar{d}'_i$ &
		$\operatorname{mean}\left(\left(d_i^{\text{max}}\right)' -
		\left(d_i^{\text{min}}\right)'\right)$ \\ \hline \hline
		\rule[-2px]{0px}{14px}
		sequential & 1.1914 & 0.0586 & 0.5069 \\
		parallel & 1.1905 & 0.0598 & 0.5076 \\
	\end{tabular}
\end{table}

Additional quality measure, which often appears in convergence and stability
proofs is quasi-uniformity~\cite{wendland2004scattered}, or in the case of
variable
nodal spacing $h$-quasi-uniformity~\cite{slak2020phd}.
Quasi-uniformity represents the notion that the relative distances between the
neighbouring nodes are approximately equal for a sequence of node sets.
Formally, it defines the \q{node set ratio} (analogous to mesh ratio) of a node
set $X$ covering $\Omega$ as a quotient of the \q{fill distance} and
\q{separation distance}:
\begin{equation}
\gamma_{X,\Omega,h} = \frac{
	\max_{p_i \in X} h_{X, \Omega}(p_i) / h(p_i)
}{
	\min_{p_i \in X} s_{X}(p_i) / h(p_i)
},
\end{equation}
where $h_{X, \Omega}(p)$ is the diameter of the largest empty ball in $\Omega$
that touches $p$, and $s_X(p)$ is the distance from $p$ to its closest
neighbour:
\begin{align}
h_{X, \Omega}(p) &= 2\,\sup_{q\, \in\, \Omega} \{
\|p-q\|, B(q, \|p-q\|) \cap X = \emptyset
\}, \\
s_{X}(p) &= \min_{q\, \in\, X \setminus\{p\}} \|p-q\|.
\end{align}
A sequence of node sets $(X_\lambda)_\lambda$ with nodal spacing functions
$h_\lambda$ is called quasi-uniform, if the node set ratios $\gamma_{X_\lambda,
	\Omega, h_\lambda}$ are uniformly bounded.

The node sets $X_i$ were generated for the 2-D clover
shape~\eqref{eq:clover} with the nodal spacing function~\eqref{eq:clover-den},
with $(\hmin)_i$ and $(\hmax)_i$ as given as $(\hmin)_i = 0.05 / 2^i$ and
$(\hmax)_i = 0.25 / 2^i$, for $i = 0, \ldots, 7$.

\begin{figure}[ht]
	\centering
	\includegraphics[height=4.5cm]{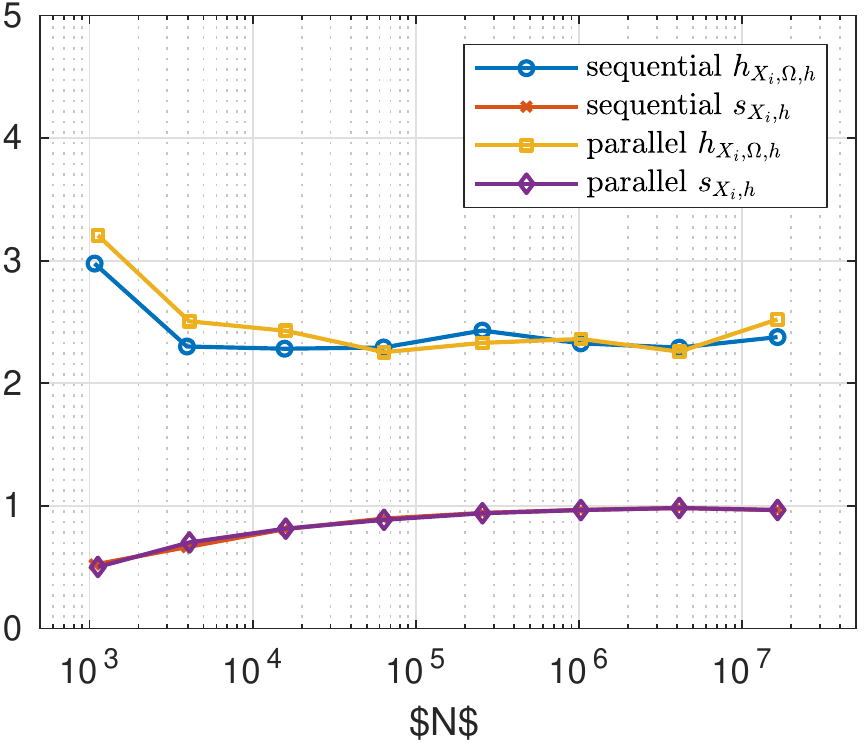}
	\includegraphics[height=4.5cm]{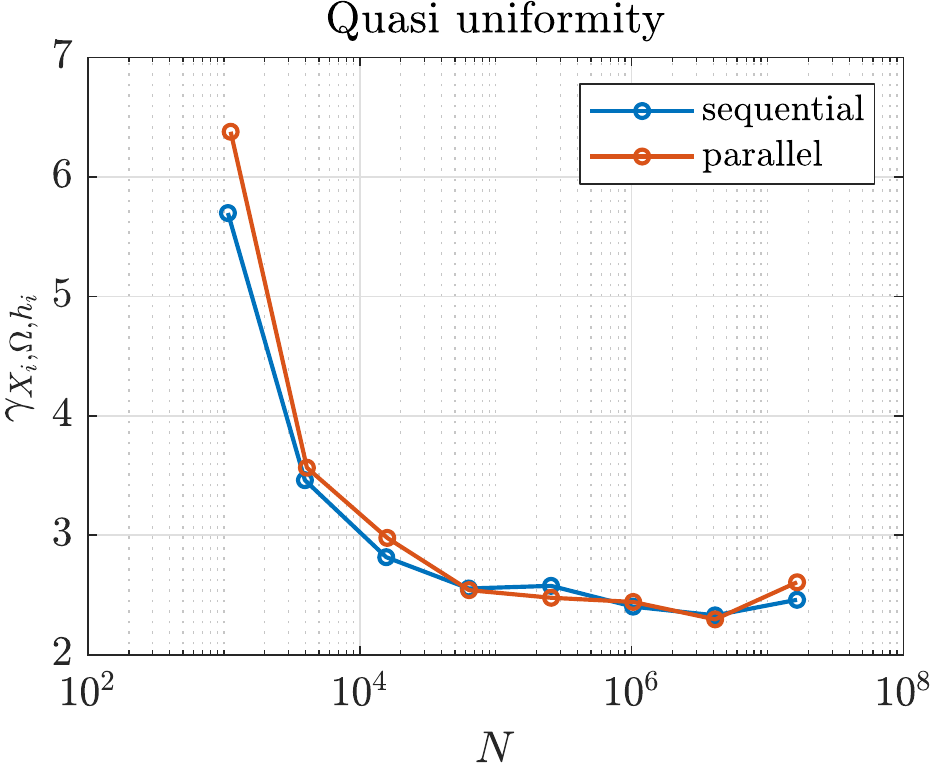}
	\caption{Quasi-uniformity analysis of the node sets generated on the 2-D
		clover domain. The left plot shows the normalized fill and separation
		distances, and the right plot shows the node set ratio.}
	\label{fig:nodes-qu}
\end{figure}

The values of $h_{X_i, \Omega, h_i} := \max_{p_j \in X_i} h_{X_i, \Omega}(p_j)
/ h_i(p_j)$
and $s_{X_i, h} := \min_{p_j \in X_i} s_{X_i}(p_j) / h_i(p_j)$ are shown in
figure~\ref{fig:nodes-qu} for the node sets $X_i$. Both values share the same
behaviour in the parallel and sequential versions of the algorithm, and it can be
deduced from the figure that the node set ratio $\gamma_{X_i, \Omega, h_i}$ is
bounded, indicating a node distribution of sufficient quality for PDE solving.

Additionally, the node distributions produced by the parallel and sequential versions of the
algorithm are compared on a model boundary value problem with a manufactured solution, to
asses any potential difference in errors or condition numbers that might appear. A Poisson problem
on the 2-D clover domain $\Omega$ is considered, specifically
\begin{align}
  \nabla^2 u(x, y) &= -\frac29 \pi^2 \sin(\pi x/3)\sin(\pi y/3) &\text{ in } \Omega, \\
  u(x, y)  &= \sin(\pi x/3)\sin(\pi y/3) &\text{ on } \partial \Omega,
\end{align}
with the solution $u_a(x, y) = \sin(\pi x/3)\sin(\pi y/3)$.

The node sets were generated for the Clover-2D domain as before with spacing $h$ as defined in
table~\ref{tab:spacing}, in such a way that
the number of
nodes ranged between 1000 and 1000000. The numerical solution was obtained using RBF-FD method
using PHS RBFs and monomial augmentation of 2nd order on stencils of 15 closest nodes.

\begin{figure}[ht]
	\centering
	\includegraphics[halfFig]{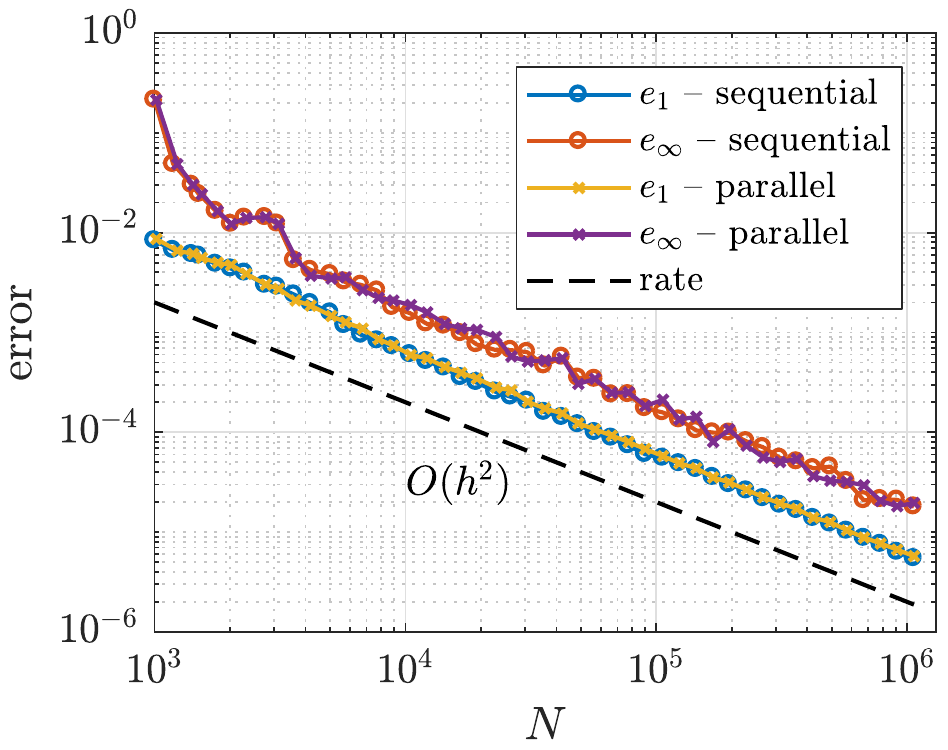}
	\includegraphics[halfFig]{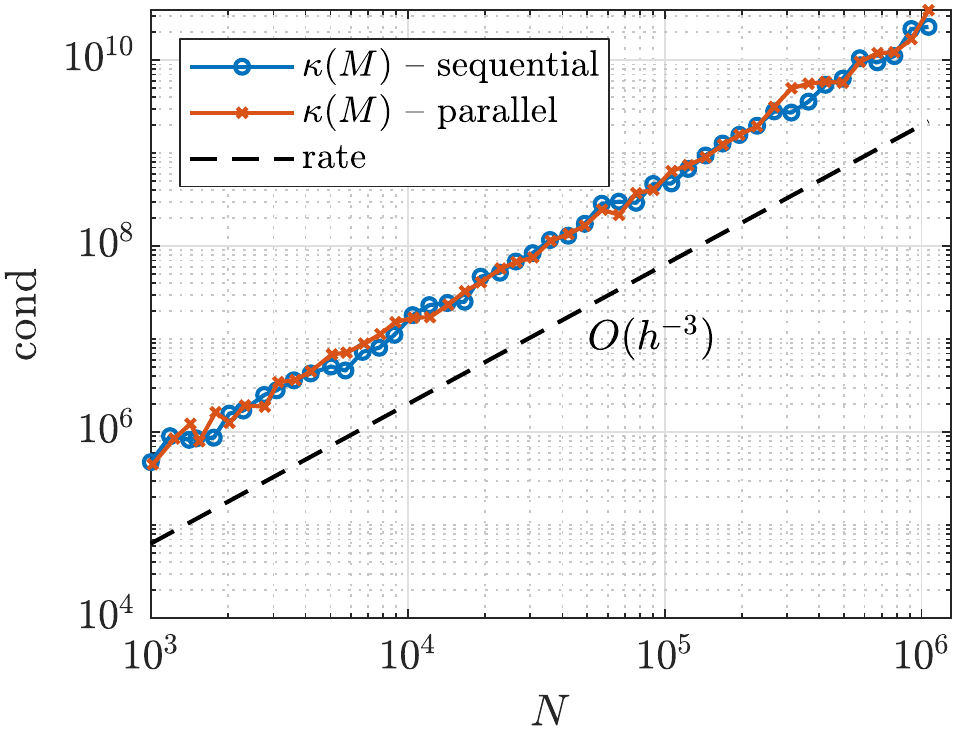}
	\caption{Comparison of estimated errors and condition numbers when solving a Poisson boundary
	value problem with nodal distributions generated by the parallel and sequential versions of the
	algorithm.}
	\label{fig:nodes-error_cond}
\end{figure}

The condition number $\kappa(M)$ of the final sparse matrix $M$ was estimated using the Matlab's
\texttt{condest} function, and the error was estimated by constructing a dense uniforms spaced grid
of nodes $G$ in the domain and comparing the difference between the numerical solution $u_h$ and
the closed-form solution $u$ at the grid nodes. The differences were aggregated in two ways,
giving estimates of the relative $L_1$, and $L_\infty$ norms:
\begin{align}
  e_1 &= \|u_h - u\|_1/\|u\|_1, \quad \|f\|_1 = \frac{1}{|G|} \sum_{p \in G} |f(p)|, \\
  e_\infty &= \|u_h - u\|_\infty/\|u\|_\infty, \quad \|f\|_\infty = \max_{p \in G} |f(p)|.
\end{align}
The results are presented in figure~\ref{fig:nodes-error_cond} along with the growth rates. No
major differences between parallel and sequential versions are observed.

\section{Discussion}
\label{sec:discussion}

In this section we discuss the some of the approaches used in the algorithm
and in the experiments.

\subsection{Parallelization overhead}

\begin{figure}[ht]
	\centering
	\includegraphics[fullFig]{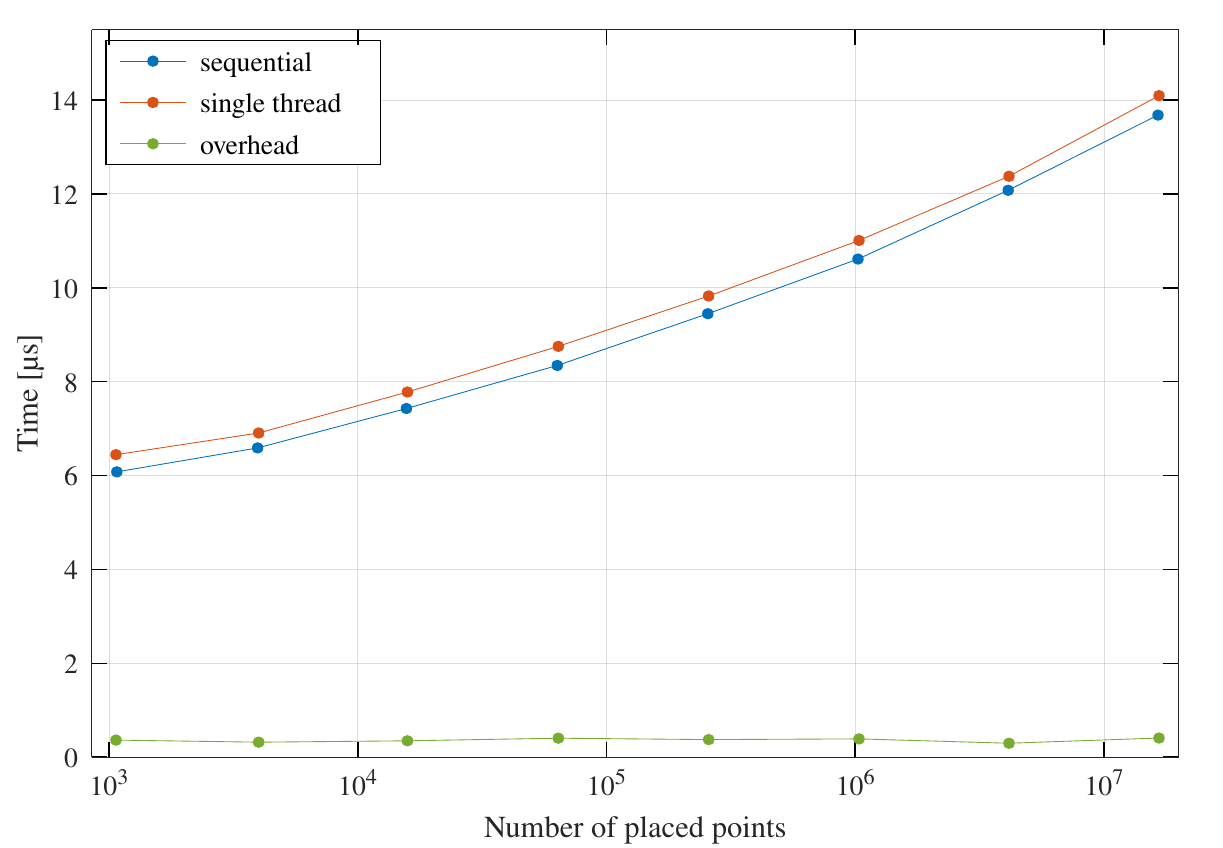}
	\caption{Comparison of the average time required for placing a single point
	between the sequential and parallel algorithms.
		Overhead is calculated as the difference between the two and comprises
		thread management and mutex ownership acquisition.}
	\label{figure:mutexOverhead}
\end{figure}

The parallel algorithm will always experience the overhead of creating, running
and managing threads, which are the features not required by the sequential
algorithm.
It also experiences the overhead of the acquiring ownership of mutexes, which
is again not required in the sequential version.
This overhead could possibly be avoided with the use of non-locking thread-safe
data structures, which is a direction for further improvements but is beyond
the scope of this paper.
These two overheads are demonstrated on the difference between the execution
times of sequential algorithm and parallel algorithm with a single worker
thread and no bootstrapping.
The differences in execution times, normalised by the number of placed points
for a clearer visualisation, are plotted in Figure~\ref{figure:mutexOverhead}.
While there are both overheads present at all experiments, the thread
management overhead should be more noticeable on the left part of the figure
where the number of placed points is low, and the mutex overhead should
dominate the right side of the figure where the number of placed points is high.
The figure shows a the overhead to be very small, with no visible trend, and
although averaged over 10 measurements, relatively noisy.
Therefore, our overall observation is that the parallel overhead on our
experimental system is about \unit[400]{ns}, which is mostly due to the mutex
acquisition, while the thread management overhead is negligible.

\subsection{Iterative method for bootstrapping}

Bootstrapping represents a method of generating a sufficient number of seed
points for the parallel algorithm to load all the worker threads.
The exact number of seed points, which should equal an integer multiple of the
number of threads for best results, cannot be achieved easily.
More than the requested number of points could be generated by bootstrapping
and then reduced down to the desired number, but such reduction would
degenerate the shape of cells induced by the seed points.
According to our preliminary testing, it seems better to leave the number of
seed points higher than requested and assign an unequal number of them to
different threads.

There are multiple viable strategies for generating the required number of seed
points, and the selection of an appropriate one could even be tailored to the
specific use case.
Below we only present a method that worked well for us and was used in all the
experiments with bootstrapping in this article.
That also means that its execution time was included in the execution times of
the parallel algorithm runs.
While the method is not optimal as it simply discards some results, which could
potentially be refined; instead, its execution time is still insignificant
compared to the whole algorithm.

The method takes the lower limit on the number of generated points as a
parameter and guarantees as many or more will be generated.
The main idea is to iterate over several values of the spacing amplification
factor (see Section~\ref{subsec:bootstrapping}), until the obtained number of
seed points is larger than the lower limit.
The proposed method starts with spacing that is intentionally too high - i.e.\
the factor $a$ is first estimated using \ref{eq:bootstrap_estimate_a}, and
multiplied by 10.
Then in a loop, the point placement algorithm is executed with the current
value for $a$, and the number of placed points is counted.
While the target number of seed points is not reached, the loop repeats with the
value of $a$ halved.
The loop iterates until a sufficient number of points is generated.
Bootstrapping is then complete, and the placed points are declared the seed
points of the parallel algorithm.

\subsection{Effects of Bootstrapping}

\begin{figure}[ht]
	\centering
	\includegraphics[fullFig]{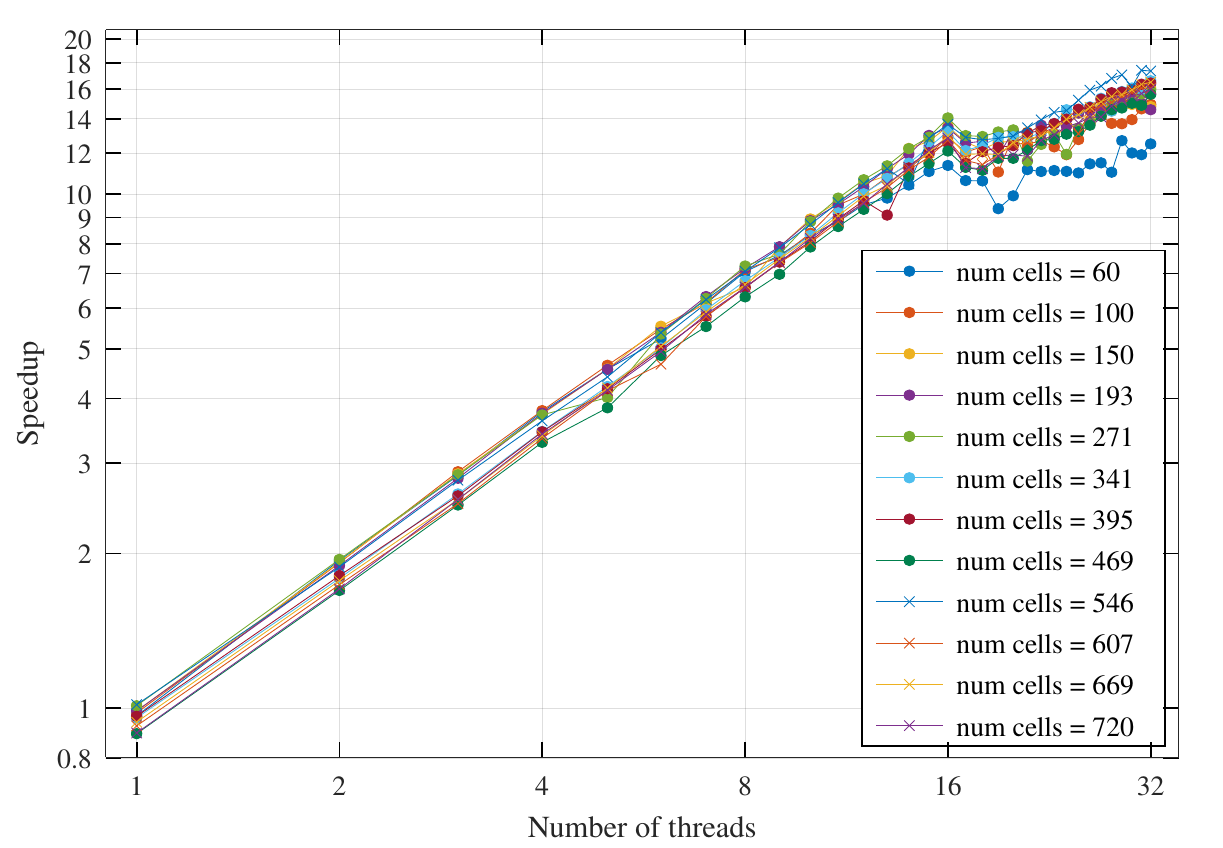}
	\caption{The speedup of the cell-based parallel algorithm as a function of
	the number of threads and the number of cells.}
	\label{figure:numSeedsInfluence}
\end{figure}

Since the sequential algorithm does not require it, bootstrapping presents an
overhead, but theoretically a relatively small one, since the total number of
seed points relative to the total number of placed points is small.
To see how significant this overhead is, we devised an experiment where we
incrementally step up the number of generated seeds to numbers far beyond the
required.
Figure~\ref{figure:numSeedsInfluence} displays how the number of seed points
influences the speedup of the presented algorithm.
It is plotted in a log--log form, for better visibility of speedups for low
number of threads.
All the experiments for this figure were executed on the clover domain, with
$n_p=4096000$, $n_c=12$, and the same random seed for all executions.
The latter ensures that even the influence of randomness is kept at a minimum.
The number of seed points was aimed at multiples (1, 2, 3, 4, 6, 8, ... 20) of
the maximum number of worker threads (32).
In the figure, the obtained number of seed points (and thus cells) is listed
since it reflects the overhead better than the requested number.
The speedup was calculated as a time fraction relative to the original
sequential algorithm, which does not perform bootstrapping but rather starts
from a single seed point, placed at coordinates $(0,0)$.

The low variation of results in the figure demonstrates the robustness of the
placement algorithm to the number of cells; at least in combination with this
particular domain and problem size.
Only the lowest numbers of cells (60) results in visibly lower speedups across
some range of experiments, while the others are clustered very closely.
No clear pattern emerges for the other numbers of cells, and while some numbers
of cells seem worse than others across the whole range, these are not the
highest numbers, which are the ones with the highest bootstrapping overhead.

Furthermore, in the case of 546 cells, parallel algorithm executed on a single
thread even outperforms the sequential algorithm.
We have ensured that these results are not an error by repeating the tests
several times.
When dividing the domain between multiple cells and performing the search on a
single thread with all the parallel algorithm's overheads, the parallel
algorithm is still faster on average.
Furthermore, whatever causes the divided domain to be filled faster, remains in
effect even when the number of threads increases; regardless of the number of
threads, the division into 546 cells results is one of the fastest executions
for a given number of threads.

\begin{figure}[ht]
	\centering
	\includegraphics[fullFig]{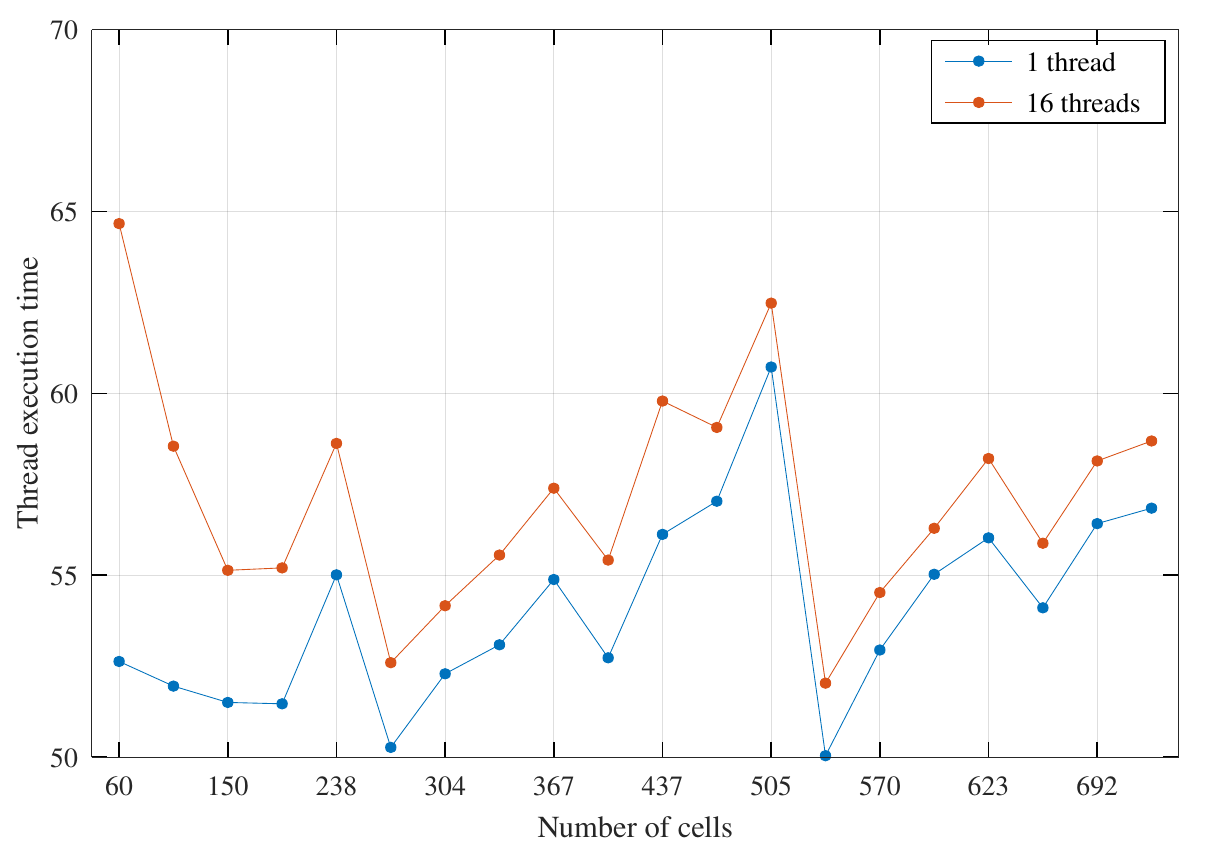}
	\caption{The single-threaded execution times contrasted to the sum of
	individual thread execution times on 16 threads.
		Markers represent the experiments while the lines are added only to aid
		the visual comparison of the variations within the two sets of
		experiments.}
	\label{figure:cellBoundariesOverhead}
\end{figure}

For a closer look at what defines the performance, we checked whether the cause
was in static load balancing.
Looking at the load balance expressed as the average ratio of idle versus work
time for threads, we saw the loads were, in general, not balanced, but there
was no correlation with the execution time.
We only noticed that the sum of execution times of individual threads varies
with the number of cells.
To get more insight into what might be the cause of varying execution times, we
show the single-thread execution time and the sum of all thread execution times
for the 16-threaded execution in Figure~\ref{figure:cellBoundariesOverhead}.
The obvious correlation between the execution times is not coincidental.
If the execution on 16-threads were compensated for the parallel overhead, the
two plotted curves would match almost perfectly.
Since we ran the experiments with a constant random seed, we obtained identical
distributions of seed points for all the tests with the same number of seed
points.
It can be reasoned then that the execution times depend mainly on the shape of
the indexing tree, governed the shape of its top level, which was in our case
uniquely prescribed by the number of seed points but in general depends on seed
point positions.
It can be reasoned then that the correlation in execution times stems form the
use of the dynamically built $k$-d tree as the spatial indexing method.
Such a tree is generally not perfectly balanced, and the extent to which it is
unbalanced in our case depends not only on chance but also on the distribution
of seed points.
Therefore, a viable option for future work is the research into other spatial
indexing methods, and how the algorithm could benefit from them.

\subsection{Points placed near cell borders}

There are limitations to cell-based parallel approach beyond the already
mentioned minimum number of seed points. %
One is also the proximity testing within individual cells.
There can always appear pairs of points from neighbouring cells that would fail
it, so some testing between neighbouring cells is required.
Neighbouring cells are all those that can theoretically interfere with the
point placement.
To eliminate the chance of such pairs appearing, two methods have been
considered:
\begin{itemize}
	\item Testing for minimal distance within the neighbouring cells while
	placing points.
	If the candidate passes the proximity test in the parent cell it then has
	to be tested also in all the neighbouring cells.
	Those tests have to be done while all the involved cells are locked for
	reading and writing, otherwise race condition could occur; %
	i.e., another thread could theoretically attempt a similar action on an own
	candidate point, both threads might find the placement non-conflicting and
	then insert, each thread into another cell, a conflicting pair of points.
	\item Inserting all the points without checks in the neighbouring cells,
	but flagging flagging points that could cause a conflict conflict in the
	neighbouring cell, e.g., adding them to a list of points
	requiring additional checks in post-processing.
	After the placement is complete, all the flagged points must be checked for
	conflicts, and the actual conflicts solved, i.e. removing one of the
	conflicting points from the domain.
	Checking must be done systematically though, to properly deal with cases where
	one point appears in more than one conflict.
	And if done in parallel, it again has to be protected against race
	conditions.
\end{itemize}

Both methods require that testing of a candidate point comprises of finding
two or more nearest cells -- one to act as the parent cell, and others to check
for possible interference with the candidate placement.
Both methods also introduce additional overheads.
We opted to use the first method for its greater simplicity but are considering
the second one as future work.

The selected method introduces the overhead of checking for neighbours in more
than one cell for all candidates on cell boundaries.
The greater the area of boundaries relative to the total domain area, the
greater the overhead of extra checking will be.
We can experimentally evaluate the overhead, by varying the number of cells,
and thus vary the ratio of cell boundary area relative to the total domain area.
Experiments can even be done on a single thread to isolate the tested parameter
form all the non-determinism of running in parallel.
The execution of algorithm on a single thread for varying number of cells was
already shown in Figure~\ref{figure:numSeedsInfluence}.
Instead of seeing a pattern of increased execution time with the increasing
area of cell borders, the experiment shows noise.
Therefore, while we still expect the overhead to be there, it seems to be
completely lost in other more significant factors to the overall execution time.
For the time being, we consider this overhead too insignificant to be worth
further analysis.

\section{Conclusion}
\label{sec:conclusion}

Parallelisation of the Poisson disc sampling algorithm for domain
discretisation is attempted by running several sequential algorithm instances on a
shared set of placed points and a shared spatial indexing method.
The instances are executed on a shared-memory machine as threads.
The main focus is put into sharing the data efficiently with minimal
synchronisation overhead.
To this end, the shared spatial indexing method (i.e.\ $k$-d tree) is
implemented with a location-aware locking mechanism.
A static load balancing mechanism is implemented through the generation of
sub-domains or cells, and the use of thread-local sets of candidate points.
Such load-balancing guides the threads to work on distant and compact
sub-domains, and decreases the chances of conflicting access to the shared data.

The implementation is experimentally tested on several irregular domains
with uniform and non-uniform nodal spacing functions.
The experiments show that the parallel algorithm is capable of high speedups.
Moreover, its performance is not influenced by the geometric complexity of the
domain.
Additionally, the experimental analysis of the presented parallelisation
approach includes the costs of various overheads and produces ideas for
future work.

The main limitation of the algorithm is in its use of dynamically constructed $k$-d tree, which can become arbitrarily unbalanced and thus cause larger than necessary search times.
Different spatial indexing strategies will be considered for replacing the $k$-d tree. There is also an array of improvements that could be made in handling border nodes. The existing border nodes should be exploited to increase the number of concurrent moving fronts and thus contribute to a more efficient parallelization. Processing the border nodes within the cellular structure, however, requires careful development and analysis to avoid extensive overheads that would make such an endeavour counterproductive.

In future, besides addressing the deficiencies mentioned above we will focus on two additional aspects. First, how to implement a distributed variant of the discussed algorithm. Second, how to utilize the presented parallel node generation algorithm to generate balanced sub-domains for the domain decomposition.

\section{Acknowledgements}

The authors would like to acknowledge the financial support of the Slovenian
Research Agency (ARRS) research core funding No. P2-0095.

\bibliographystyle{elsarticle-num}

\end{document}